\RequirePackage{fix-cm}
\documentclass[twocolumn,epjc3]{svjour3}
\smartqed  
\usepackage{amsmath}
\usepackage{amssymb}
\usepackage{amssymb,bbold}
\usepackage{kantlipsum,widetext}
\usepackage{subcaption}

\RequirePackage{graphicx}
\RequirePackage[colorlinks,citecolor=blue,urlcolor=blue,linkcolor=blue]{hyperref}
\RequirePackage[numbers,sort&compress]{natbib}

\usepackage{xcolor}

\journalname{Eur. Phys. J. C}
\begin{document}

\title{Scalar and vector bosons in a Bonnor-Melvin-$\Lambda$ spacetime: an exact Duffin-Kemmer-Petiau analysis
}


\author{Francisco A. Cruz Neto\thanksref{addr2} \and Luis B. Castro\thanksref{e1,addr2,addr1}
}

\thankstext{e1}{e-mail: lrb.castro@ufma.br}


\institute{Programa de P\'{o}s-graduação em F\'{i}sica, Universidade Federal do Maranh\~{a}o, Campus Universit\'{a}rio do Bacanga, 65080-805, S\~{a}o Lu\'{\i}s, Maranh\~{a}o, Brazil. \label{addr2} \and
Coordenaç\~{a}o do Curso de F\'{i}sica - Bacharelado, Universidade Federal do Maranh\~{a}o, Campus Universit\'{a}rio do Bacanga, 65080-805, S\~{a}o Lu\'{\i}s, Maranh\~{a}o, Brazil. \label{addr1}
}

\date{Received: date / Accepted: date}

\maketitle

\begin{abstract}
We study scalar and vector bosons in the Bonnor--Melvin--$\Lambda$ spacetime within the Duffin--Kemmer--Petiau (DKP) formalism. By employing Umezawa's projection operators, we separate the physical spin-0 and spin-1 sectors and derive the corresponding exact second-order equations in the full curved spacetime, without relying on the conical approximation. For the scalar sector, the radial equation reduces to a Schr\"odinger-like equation with a trigonometric P\"oschl--Teller effective potential. In the vector sector, the longitudinal mode is governed by the same effective potential, whereas the transverse polarizations are described by generalized trigonometric P\"oschl--Teller potentials. Because the me\-tric function vanishes at a discrete set of radial points, the radial dynamics is naturally formulated as a singular Sturm--Liouville problem on a fundamental interval, with the physical radial domain fixed by the Friedrichs self-adjoint extension of the corresponding singular radial operators. As a result, all physical sectors exhibit purely discrete radial spectra, and their eigenfunctions are obtained in closed form. These results provide a unified exact treatment of scalar and vector bosons in the Bonnor--Melvin--$\Lambda$ spacetime, complement previous analyses based on the conical approximation, and clarify the role of the global geometric structure of the background in shaping confinement and spectral pro\-perties. 

\end{abstract}

\section{Introduction}
\label{intro}

The first-order Duffin--Kemmer--Petiau (DKP) formalism provides a unified framework for the description of spin-0 and spin-1 bosons \cite{Petiau1936,Kemmer1938,PR54:1114:1938,Kemmer1939}. It has long been regarded as an alternative to the Klein--Gordon and Proca theories, particularly because, although these formulations are equivalent in the free case and for minimally coupled vector interactions \cite{PLA244:329:1998,PLA268:165:2000,PRA90:022101:2014}, the DKP theory admits a richer coupling structure that cannot, in general, be implemented within the conventional second-order approaches \cite{PRD15:1518:1977,JPA12:665:1979}. For this reason, the DKP equation has been extensively employed in the investigation of relativistic bosonic systems under a variety of external interactions and geometric backgrounds.

In recent years, relativistic wave equations in curved space-times have received considerable attention, since they offer a useful framework for probing the interplay between gravitation, topology and quantum dynamics. Within this context, several studies have addressed the DKP equation in nontrivial geometries, showing that curvature and global properties of the background may significantly affect the spectral structure, localization properties and transport of bosonic fields \cite{EPJC44:287:2005,IJMPA25:2747:2010,AP343:40:2014,EPJP130:236:2015,EPJC75:287:2015,EPJC76:61:2016,EPJP132:541:2017,EPL118:10002:2017,
EPJC78:93:2018,GRG50:104:2018,IJMPA34:1950056:2019,IJMPA34:1950082:2019,GRG52:25:2020,IJMPA35:2050107:2020,MPLA36:2150059:2021,IJMPE30:2150050:2021,
CQG39:075007:2022,PS98:065224:2023,FBS64:13:2023}. Such analyses are relevant not only from the mathematical point of view, but also because they provide insights into the behavior of relativistic particles in strong gravitational or cosmological environments.

Magnetic fields also play a central role in a wide range of astrophysical and cosmological scenarios. Their influence extends from compact objects and accretion disks to galactic and intergalactic scales, and their incorporation into the space-time geometry leads to nontrivial Einstein--Maxwell configurations. Among the best-known examples is the Bonnor--Melvin solution, which describes a cylindrically symmetric magnetic universe with a magnetic field aligned along the symmetry axis \cite{PPSA67:225:1954,PL8:65:1964}. More recently, this solution was generalized to include a nonvanishing cosmological constant, giving rise to the Bonnor--Melvin--$\Lambda$ spacetime \cite{PRD99:044058:2019}. In this generalized setting, the geometry remains static and cylindrically symmetric, while the intrinsic magnetic field acquires a periodic structure controlled by the cosmological constant. In particular, the metric function is proportional to $\sin(\sqrt{2\Lambda}\,r)$, which implies that the background is not asymptotically flat and naturally introduces a sequence of singular points separating neighboring radial cells. 

The quantum dynamics of particles in Bonnor--Melvin--$\Lambda$ backgrounds has attracted increasing attention. Fermi\-onic and scalar sectors have already been discussed in related backgrounds \cite{IJMPA39:2450032:2024,FBS66:7:2024,PS100:035302:2025,NPB1004:116569:2024,GRG57:111:2025}, and charged scalar bosons were recently investigated within the DKP formalism in the presence of an additional vector potential \cite{EPJC84:536:2024}. In that analysis, the complexity of the exact radial equation led to the adoption of the conical approximation $\sqrt{2\Lambda}\,r\ll 1$, or equivalently $\Lambda\ll 1$, under which the problem reduces to effective Coulomb-like and oscillator-like systems depending on the form of the vector potential. Although this approximation is well suited for probing microscopic regimes, it also hides an important feature of the exact Bonnor--Melvin--$\Lambda$ geometry, namely the periodic structure generated by the trigonometric metric function.

More recently, the global and singular structure of the
Bonnor--Melvin--\(\Lambda\) geometry has been analyzed in detail in
several related contexts~\cite{EPJC85:509:2025,PLB866:139569:2025,NPB1016:116920:2025,NPB1018:116998:2025,IJGMMP0:2650065:2025,IJGMMP0:2650145:2025}. These studies have shown that the
zeros of the metric function play a decisive role in the definition of
the admissible radial domains, since they divide the spacetime into
finite radial sectors and may be interpreted as geometrically induced
boundaries or domain-wall-like structures. In particular, the
corresponding wave equations are not naturally defined as ordinary
scattering problems on the half-line, but rather as singular
boundary-value problems on finite radial cells. These results provide
the appropriate geometrical background for the present analysis.

Motivated by this cell-like structure, we investigate the dynamics of scalar and vector bosons in the exact
Bonnor--Melvin--\(\Lambda\) geometry within the projected DKP formalism. In contrast with treatments based
on the conical approximation, the present approach keeps the full
trigonometric dependence of the metric function and therefore allows us to determine how the finite radial cells modify the effective
potentials, endpoint conditions, and spectra of the spin-0 and spin-1
sectors.

Another motivation for the present work comes from the spin-1 sector itself. While the scalar sector of the DKP theory has been studied in a number of curved backgrounds, the corresponding vector sector has also been formulated systematically by means of Ume\-za\-wa's projection operators, which lead to the Proca equation in curved spacetime \cite{LUNARDI2001}. Here, we take a further step by rewriting that equation, through commutators of covariant derivatives, in a form in which the Ricci tensor appears explicitly. This provides the natural starting point for the analysis of the physical vector components in general curved spacetimes, and in particular for the exact treatment developed here in the Bonnor--Melvin--$\Lambda$ spacetime. The same philosophy may also be applied to the spin-0 sector, allowing one to treat both scalar and vector bosons within a unified projected DKP framework.

In this work, we investigate scalar and vector bosons in the Bonnor--Melvin--$\Lambda$ spacetime within a unified projected Duffin--Kemmer--Petiau framework. Our main goal is to determine how the exact trigonometric structure of the background, rather than its conical approximation, modifies the radial dynamics, admissibility conditions, and spectral properties of the physical spin-0 and spin-1 sectors. To this end, we derive the exact projected second-order equations directly in the full geometry and show that the resulting radial problems are analytically solvable on a fundamental interval bounded by the zeros of the metric function. This allows us to identify the geometric origin of the discrete spectra and to clarify the different roles played by the scalar, longitudinal, and transverse vector modes in the exact Bonnor--Melvin--$\Lambda$ background.

This paper is organized as follows. In Sect.~\ref{sec:projected_dkp}, we briefly review the DKP equation in curved space-time and introduce Umezawa's projection operators for the spin-0 and spin-1 sectors. In Sect.~\ref{sec:bm_geometry}, we summarize the Bonnor--Melvin--$\Lambda$ geometry and establish the basic geometric quantities required in the DKP formalism. In Sect.~\ref{sec:scalar_sector}, we analyze the scalar sector and derive the exact radial equation, its Schr\"odinger-like form, the appropriate boundary conditions and the corresponding exact spectrum. In Sect.~\ref{sec:vector_sector}, we turn to the vector sector, derive the equations for the physical polarizations and discuss their exact analytical solutions. Finally, in Sect.~\ref{sec:conclusions}, we present our conclusions.                  

\section{Projected spin-0 and spin-1 sectors of the DKP equation}
\label{sec:projected_dkp}

We begin with the free Duffin--Kemmer--Petiau equation in a curved spacetime \cite{GRG34:491:2002,GRG34:1941:2002,EPJC75:287:2015,EPJC76:61:2016,EPJC84:536:2024} ($\hbar=c=1$),
\begin{equation}
\left(i\beta^\mu \nabla_\mu - M\right)\Psi=0,
\label{eq:DKP_curved_free}
\end{equation}
where the curved-space matrices $\beta^\mu=e^\mu{}_{\bar a}\beta^{\bar{a}}$ satisfy the DKP algebra and
\begin{equation}
\nabla_\mu=\partial_\mu-\Gamma_\mu
\label{eq:cov_derivative}
\end{equation}
is the covariant derivative. The spinorial connection is defined by
\begin{equation}
\Gamma_{\mu}=\frac{1}{2}\omega_{\mu\bar{a}\bar{b}}[\beta^{\bar{a}},\beta^{\bar{b}}],
\label{eq:spinorial_connection}
\end{equation}
where the spin connection coefficients are given by
\begin{equation}
\omega_{\mu}{}^{\bar{a}\bar{b}}=e_\alpha{}^{\bar a}e^\nu{}^{\bar{b}}\Gamma^{\alpha}_{\mu\nu}-e^\nu{}^{\bar{b}}\partial_{\mu}e_\nu{}^{\bar{a}},
\label{eq:spin_connection_coefficients}
\end{equation} 
with $\omega_{\mu}{}^{\bar{a}\bar{b}}=-\omega_{\mu}{}^{\bar{b}\bar{a}}$ and the Christoffel symbols are given by
\begin{equation}
\Gamma_{\mu\nu}^{\alpha}=\frac{g^{\alpha\lambda}}{2}\left(\partial_{\mu}g_{\lambda\nu}+\partial_{\nu}g_{\lambda\mu}-\partial_{\lambda}g_{\mu\nu}\right).
\label{eq:christoffel_symbols}
\end{equation}

The tetrads $e_\mu{}^{\bar a}$ satisfy
\begin{eqnarray}
\eta^{\bar{a}\bar{b}} &=& e_\mu{}^{\bar a}e_\nu{}^{\bar b}g^{\mu\nu},\label{eq:tetrad_relation_1}\\
g_{\mu\nu} &=& e_\mu{}^{\bar a}e_\nu{}^{\bar b}\eta_{\bar{a}\bar{b}},\label{eq:tetrad_relation_2}
\end{eqnarray}
and
\begin{equation}
e_\mu{}^{\bar a}e^\nu{}_{\bar a}=\delta_\mu^\nu, \qquad
e^\mu{}_{\bar a}e_\mu{}^{\bar b}=\delta_{\bar a}^{\bar b}.
\label{eq:tetrad_inverse_relations}
\end{equation}
Latin indices are raised and lowered by the Minkowski metric tensor $\eta^{\bar{a}\bar{b}}$ with signature $(+,-,-,-)$, whereas Greek ones are raised and lowered by the spacetime metric $g^{\mu\nu}$.

The usual four-current is given by
\begin{equation}
J^{\mu}=\frac{1}{2}\bar{\Psi}\beta^{\mu}\Psi,
\end{equation}
where the adjoint spinor is defined by $\bar{\Psi}=\Psi^{\dagger}\eta^{0}$ with $\eta^{0}=2\beta^{0}\beta^{0}-1$, in such a way that $(\eta^{0}\beta^{\mu})^{\dagger}=\eta^{0}\beta^{\mu}$, that is, the matrices $\beta^{\mu}$ are Hermitian with respect to $\eta^{0}$.

The aim of this section is to isolate the physical
spin-0 and spin-1 sectors by means of Umezawa's projection operators and to
derive the corresponding second-order equations that will be analyzed in the
Bonnor--Melvin--$\Lambda$ background in the following sections.

\subsection{Projection onto the spin-0 sector}
\label{subsec:spin0_projection}

To project the scalar sector, we introduce the standard Umezawa operators \cite{UMEZAWA1956}
\begin{equation}
P=-(\beta^{0})^{2}(\beta^{1})^{2}(\beta^{2})^{2}(\beta^{3})^{2},
\label{eq:scalar_projectors}
\end{equation}
which satisfies $P^{2}=P$, $P^\mu=P\beta^\mu$ and $^{\nu}\!P=(P^{\nu})^{\dagger}=\beta^{\nu}P$. These operators isolate the physical scalar degree of freedom and its auxiliary
vector component. In particular, they satisfy the relations
\begin{equation}
P^\mu\beta^\nu=P\,g^{\mu\nu},
\qquad
P S^{\mu\nu}=0,
\label{eq:scalar_proj_relations}
\end{equation}
where $S^{\mu\nu}=[\beta^\mu,\beta^\nu]$.

Since $P$ transforms as a scalar and $P^\mu$ as a vector, one has
\begin{equation}
P\nabla_\mu\Psi=\nabla_\mu(P\Psi),
\qquad
P^\nu\nabla_\mu\Psi=\nabla_\mu(P^\nu\Psi).
\label{eq:scalar_proj_cov}
\end{equation}
Applying $P$ and $P^\nu$ to Eq.~\eqref{eq:DKP_curved_free}, we obtain
\begin{equation}
i\nabla_\mu(P^\mu\Psi)-M(P\Psi)=0,
\label{eq:P_projection}
\end{equation}
and
\begin{equation}
i\nabla^\nu(P\Psi)-M(P^\nu\Psi)=0.
\label{eq:Pmu_projection}
\end{equation}
Defining the physical scalar field by
\begin{equation}
\varphi \equiv P\Psi,
\label{eq:phi_definition}
\end{equation}
Eq.~\eqref{eq:Pmu_projection} yields
\begin{equation}
P^\mu\Psi=\frac{i}{M}\nabla^\mu\varphi.
\label{eq:scalar_auxiliary}
\end{equation}
Substituting Eq.~\eqref{eq:scalar_auxiliary} into Eq.~\eqref{eq:P_projection}, we
find
\begin{equation}
\left(\nabla_\mu\nabla^\mu+M^2\right)\varphi=0.
\label{eq:KG_projected}
\end{equation}
Therefore, the projected spin-0 sector of the DKP equation is described by a
second-order Klein--Gordon-type equation for the physical component
$\varphi=P\Psi$, while the remaining projected components are auxiliary and are
fully determined by Eq.~\eqref{eq:scalar_auxiliary}.

We now turn back to the DKP current. The $P$-algebra summarized in \ref{A:1} implies that%
\begin{eqnarray}
J^{\mu } &=&\frac{1}{2}\,\bar{\Psi}\left( P^{\mu }+\,^{\mu }\!P\right) \Psi
\notag \\
&=&\frac{i}{2M}\left[
\left(\bar{\Psi}P\right)\partial ^{\mu}\left(P\Psi\right)
-\left(P\Psi\right)\,\partial^{\mu}\left(\bar{\Psi}P\right)
\right].
\label{eq:projected_scalar_current}
\end{eqnarray}
Using the projected field \eqref{eq:phi_definition}, the above expression becomes
\begin{equation}
J^{\mu}=\frac{i}{2M}\left(
\varphi^{\ast}\partial ^{\mu}\varphi
-\varphi\,\partial ^{\mu}\varphi^{\ast}
\right).
\label{eq:projected_scalar_current2}
\end{equation}
This is precisely the Klein--Gordon current. In other words, the DKP current in
the projected spin-0 sector is equivalent to the Klein--Gordon current, as
reported in \cite{PRA90:022101:2014}.

\subsection{Projection onto the spin-1 sector}
\label{subsec:spin1_projection}

To isolate the vector sector, we employ Umezawa's projectors
\begin{equation}
R^\mu=e^\mu{}_{\bar a}R^{\bar a},
\qquad
R^{\mu\nu}=e^\mu{}_{\bar a}e^\nu{}_{\bar b}R^{\bar a\bar b},
\label{eq:vector_projectors}
\end{equation}
where
\begin{equation}
R^{\bar a}\equiv
(\beta^{\bar 1})^2(\beta^{\bar 2})^2(\beta^{\bar 3})^2
\left(\beta^{\bar a}\beta^{\bar 0}-\eta^{\bar a\bar 0}\right),
\quad
R^{\bar a\bar b}=R^{\bar a}\beta^{\bar b}.
\label{eq:vector_projectors_flat}
\end{equation}
These operators satisfy the identities
\begin{equation}
R^{\mu\nu}=-R^{\nu\mu},
\qquad
R^{\mu\nu}\beta^\alpha=g^{\nu\alpha}R^\mu-g^{\mu\alpha}R^\nu,
\label{eq:vector_proj_rel1}
\end{equation}
and
\begin{equation}
R^\mu S^{\nu\alpha}=g^{\mu\nu}R^\alpha-g^{\mu\alpha}R^\nu.
\label{eq:vector_proj_rel2}
\end{equation}
Moreover, the projected quantities satisfy
\begin{equation}
R^\nu\nabla_\mu\Psi=\nabla_\mu(R^\nu\Psi),
\qquad
R^{\alpha\nu}\nabla_\mu\Psi=\nabla_\mu(R^{\alpha\nu}\Psi).
\label{eq:vector_proj_cov}
\end{equation}

Applying $R^\nu$ and $R^{\mu\nu}$ to Eq.~\eqref{eq:DKP_curved_free}, one obtains
\begin{equation}
R^\nu\Psi=\frac{i}{M}\nabla_\mu(R^{\nu\mu}\Psi),
\label{eq:Rnu_projected}
\end{equation}
and
\begin{equation}
R^{\nu\mu}\Psi=\frac{i}{M}U^{\mu\nu},
\label{eq:Rmunu_projected}
\end{equation}
where
\begin{equation}
U^{\mu\nu}=\nabla^\mu(R^\nu\Psi)-\nabla^\nu(R^\mu\Psi).
\label{eq:U_definition}
\end{equation}
Defining the physical vector field as
\begin{equation}
W^\mu \equiv R^\mu\Psi,
\label{eq:W_definition}
\end{equation}
Eqs.~\eqref{eq:Rnu_projected} and \eqref{eq:Rmunu_projected} combine into
\begin{equation}
\nabla_\mu U^{\mu\nu}+M^2W^\nu=0,
\label{eq:Proca_curved}
\end{equation}
which is the Proca equation in curved spacetime \cite{LUNARDI2001}.

To obtain a second-order equation for each physical component, we use the
commutators of covariant derivatives acting on vectors and rank-2 tensors.
Then Eq.~\eqref{eq:Proca_curved} can be rewritten as
\begin{equation}
\left(\nabla_{\mu}\nabla^{\mu}+M^2\right)W^{\nu}
- \mathcal{R}_\lambda{}^{\nu} W^\lambda
-\nabla^{\nu}\!\left(\nabla_{\mu} W^{\mu}\right)=0,
\label{eq:spin1_second_order}
\end{equation}
together with the subsidiary relation
\begin{equation}
\nabla_\mu W^\mu=\frac{1}{M^2}\mathcal{R}_{\alpha\lambda}U^{\alpha\lambda},
\label{eq:spin1_subsidiary}
\end{equation}
where $\mathcal{R}_{\mu\nu}$ denotes the Ricci tensor. Eq.~\eqref{eq:spin1_subsidiary} is formally reminiscent of the so-called anomalous term that appears in the structure of the electromagnetic interaction in spin-1 DKP theory in flat spacetime \cite{EPL146:40001:2024}. In this sense, one may view the curvature-dependent contribution on the right-hand side as a gravitational analogue at the level of the subsidiary relation.

For a torsionless Levi--Civita connection, $\mathcal{R}_{\alpha\lambda}$ is symmetric,
whereas $U^{\alpha\lambda}$ is antisymmetric. Hence,
\begin{equation}
\mathcal{R}_{\alpha\lambda}U^{\alpha\lambda}=0,
\end{equation}
and Eq.~\eqref{eq:spin1_subsidiary} reduces to the transversality condition
\begin{equation}
\nabla_\mu W^\mu=0.
\label{eq:transversality_condition}
\end{equation}
Therefore, the second-order wave equation for the physical vector field becomes
\begin{equation}
\left(\nabla_\mu\nabla^\mu+M^2\right)W^\nu
- \mathcal{R}_\lambda{}^\nu W^\lambda=0.
\label{eq:spin1_second_order_reduced}
\end{equation}

We now return to the DKP current. Using the spin-1 subalgebra given in
\ref{A:2}, and keeping the notation of Ref.~\cite{JMP14:1760:1973},
where explicit summations are used instead of the Einstein convention, one finds
\begin{eqnarray}
J^{\mu } &=&\frac{1}{2}\sum_{\lambda }\bar{\Psi}
\left( ^{\mu \lambda }\!\,V^{\lambda}
+\,^{\lambda }\!\,V^{\lambda \mu }\right) \Psi
\notag \\
&=&\frac{1}{2}\sum_{\lambda }\bar{\Psi}\left( ^{\lambda }\!R\right)
\left( R^{\lambda\mu}\Psi \right)
+\frac{1}{2}\sum_{\lambda }\bar{\Psi}\left( ^{\mu \lambda }\!R\right)
\left( R^{\lambda }\Psi \right)
\notag \\
&=&\frac{i}{2M}\sum_{\lambda }\left[
\bar{\Psi}\left( ^{\lambda}\!R\right) U^{\mu \lambda}
-\left(R^{\lambda }\Psi \right)
\left( U^{\mu \lambda }\right)^{\ast }
\right].
\label{eq:vector_current_sum}
\end{eqnarray}
Using the projected field \eqref{eq:W_definition}, the above expression can be rewritten in the compact modern form
\begin{equation}
J^\mu
=
-\frac{i}{2M}
\left(
W_\lambda^{\ast}U^{\mu\lambda}
-
W_\lambda\,U^{\mu\lambda\ast}
\right),
\label{eq:projected_vector_current}
\end{equation}
which is precisely the Proca current for a complex vector field \cite{PRA90:022101:2014}. For the
Levi--Civita connection, the antisymmetric tensor may also be written as
\begin{equation}
U_{\mu\nu}
=
\nabla_\mu W_\nu-\nabla_\nu W_\mu
=
\partial_\mu W_\nu-\partial_\nu W_\mu,
\label{eq:antisymmetric_tensor_partial}
\end{equation}
whereas
\begin{equation}
U^{\mu\nu}=g^{\mu\alpha}g^{\nu\beta}U_{\alpha\beta}.
\label{eq:antisymmetric_tensor_raised}
\end{equation}

Equations \eqref{eq:KG_projected} and \eqref{eq:spin1_second_order_reduced}
show that the DKP equation naturally yields second-order wave equations for the
physical scalar and vector sectors once the appropriate projection operators are
introduced. In the scalar case, the dynamics is governed by a Klein--Gordon-type
equation for the projected field $\varphi=P\Psi$, and the corresponding DKP current reduces to the usual Klein--Gordon current. In the vector case, the projected field $W^{\mu}=R^{\mu}\Psi$ satisfies a Proca-type equation in which the
Ricci tensor appears explicitly, while the projected DKP current is equivalent to the Proca current for a complex vector field. 

It is important to stress, however, that the projected DKP formalism is
not merely a rederivation of the Klein--Gordon and Proca equations. In
the minimally coupled case considered here, the projected spin-0 and
spin-1 sectors indeed reduce to Klein--Gordon- and Proca-type
dynamics. Nevertheless, the DKP framework provides a unified
first-order description of scalar and vector bosons, in which the
physical components, auxiliary components, and conserved currents are
obtained from the same algebraic structure. This is particularly useful
in curved spacetime, where the tetrad, spin connection, and projection
operators determine in a systematic way how the geometry enters each
physical sector.

Moreover, consistently with the motivation outlined in the Introduction, the DKP equation admits a richer coupling structure than the
usual second-order Klein--Gordon and Proca formulations~\cite{PRD15:1518:1977,JPA12:665:1979}. Although such
nonminimal interactions are not introduced in the present work, the
projected approach adopted here provides the appropriate starting point
for including them in future extensions while keeping a direct
connection with the corresponding Klein--Gordon and Proca limits.
These features provide the natural motivation for the exact analysis in
the Bonnor--Melvin--\(\Lambda\) spacetime, to which we now turn.

\section{Bonnor--Melvin--$\Lambda$ spacetime and geometric structure}
\label{sec:bm_geometry}

We now specialize the general projected formalism developed in the previous section to the Bonnor--Melvin--$\Lambda$ spacetime. This background is a static and cylindrically symmetric solution of the Einstein--Maxwell equations with nonvanishing cosmological constant, and it is described by the line element
\begin{equation}
ds^2=dt^2-dr^2-S^2(r)\,d\phi^2-dz^2,
\label{eq:BM_metric_compact}
\end{equation}
in cylindrical coordinates $(t,r,\phi,z)$, with $-\infty<z<\infty$, $r\geq 0$ and $0\leq\phi\leq 2\pi$. The metric function is defined as
\begin{equation}
S(r)=\sigma\sin(ar),
\label{eq:def_a_S}
\end{equation}
where $a=\sqrt{2\Lambda}$, $\sigma$ is an integration constant, and $\Lambda$ is the cosmological constant. For this geometry, the Ricci scalar is constant,
\begin{equation}
\mathcal{R}=-4\Lambda,
\label{eq:Ricci_scalar}
\end{equation}
showing that the spacetime is not asymptotically flat. Furthermore, the background supports an intrinsic magnetic field aligned with the symmetry axis,
\begin{equation}
H(r)=\sqrt{\Lambda}\,S(r),
\label{eq:intrinsic_magnetic_field}
\end{equation}
whose periodic structure is controlled by $\Lambda$.

In this notation, the zeros of $S(r)$ occur at
\begin{equation}
r_n=\frac{n\pi}{a},
\qquad
n=0,1,2,\dots,
\label{eq:zeros_S}
\end{equation}
and naturally divide the radial domain into fundamental intervals. This trigonometric structure will play a central role in the spectral analysis of both the scalar and vector sectors.

A convenient diagonal tetrad adapted to the line element \eqref{eq:BM_metric_compact} is
\begin{equation}
e^\mu{}_{\bar a}=
\begin{pmatrix}
1 & 0 & 0 & 0\\
0 & 1 & 0 & 0\\
0 & 0 & \dfrac{1}{S(r)} & 0\\
0 & 0 & 0 & 1
\end{pmatrix},
\qquad
e_\mu{}^{\bar a}=
\begin{pmatrix}
1 & 0 & 0 & 0\\
0 & 1 & 0 & 0\\
0 & 0 & S(r) & 0\\
0 & 0 & 0 & 1
\end{pmatrix}.
\label{eq:tetrad_BM}
\end{equation}
Accordingly, the curved-space DKP matrices are
\begin{align}
\beta^{t} &= \beta^{\bar{0}},\\
\beta^{r} &= \beta^{\bar{1}},\\
\beta^{\phi} &= \frac{1}{S(r)}\beta^{\bar{2}},\\
\beta^{z} &= \beta^{\bar{3}}.
\label{eq:beta_curved_BM}
\end{align}
For this choice of tetrad, the only nonvanishing spinorial connection is the azimuthal one,
\begin{equation}
\Gamma_\phi=-a\,\sigma\cos(ar)\,[\beta^{\bar 1},\beta^{\bar 2}],
\label{eq:spin_connection_phi}
\end{equation}
so that the covariant derivatives become
\begin{align}
\nabla_t &= \partial_t,\\
\nabla_r &= \partial_r,\\
\nabla_\phi &= \partial_\phi+a\,\sigma\cos(ar)\,[\beta^{\bar 1},\beta^{\bar 2}],\\
\nabla_z &= \partial_z.
\label{eq:cov_derivatives_BM}
\end{align}
Moreover, the curved-space beta matrices are covariantly constant $\nabla_\mu\beta^\mu=0$, which guarantees the conservation of the DKP current in this background \cite{EPJC84:536:2024}.

The Christoffel symbols associated with the metric \eqref{eq:BM_metric_compact} are particularly simple. The only nonvanishing independent components are
\begin{align}
\Gamma^{r}_{\phi\phi} &= -S(r)S^{\prime}(r),\\
\Gamma^{\phi}_{r\phi} &= \Gamma^{\phi}_{\phi r}=\frac{S^{\prime}(r)}{S(r)}=a\cot(ar).
\label{eq:Christoffel_BM}
\end{align}
From these expressions, the nonvanishing components of the Ricci tensor are
\begin{align}
\mathcal{R}_{rr} &= -\frac{S^{\prime\prime}(r)}{S(r)}=a^2=2\Lambda,\\
\mathcal{R}_{\phi\phi} &= -S(r)S^{\prime\prime}(r)=a^2S^2(r)=2\Lambda\,S^2(r),
\label{eq:Ricci_tensor_BM}
\end{align}
whereas
\begin{equation}
\mathcal{R}_{tt}=\mathcal{R}_{zz}=0.
\label{eq:Ricci_tensor_zero}
\end{equation}
These results are fully consistent with Eq.~\eqref{eq:Ricci_scalar}, since
\begin{equation}
\mathcal{R}=g^{\mu\nu}\mathcal{R}_{\mu\nu}=-2a^2=-4\Lambda.
\label{eq:Ricci_check}
\end{equation}

The structure of the Ricci tensor is especially important for the spin-1 sector. Indeed, after projection, the vector field satisfies a Proca-type equation with an explicit curvature contribution proportional to $\mathcal{R}_\nu{}^\lambda$. In the present background, this term acts only on the radial and azimuthal components of the physical vector field, while the temporal and longitudinal directions remain unaffected by the Ricci tensor. This fact allows a natural separation between longitudinal and transverse polarizations in the exact vector analysis. On the other hand, for the spin-0 sector, the geometric information enters through the Laplace--Beltrami operator built from the metric \eqref{eq:BM_metric_compact}, which leads to the exact radial equation analyzed in the following section. 

It is worth emphasizing that the periodicity of the metric function
\(S(r)=\sigma\sin(ar)\) makes the exact Bonnor--Melvin--\(\Lambda\)
geometry qualitatively different from its conical approximation. The relevance of the zeros of this metric function for the definition of the admissible radial domains has been emphasized in Refs.~\cite{EPJC85:509:2025,PLB866:139569:2025,NPB1016:116920:2025,NPB1018:116998:2025,IJGMMP0:2650065:2025,IJGMMP0:2650145:2025}. According to Eq.~\eqref{eq:zeros_S}, these zeros occur at \(r_n=n\pi/a\). At these points the angular orbit collapses, the
spatial volume element \(\vert S(r)\vert\,dr\,d\phi\,dz\) degenerates, and the inverse
metric component \(g^{\phi\phi}=-1/S^2(r)\) diverges. For modes with
azimuthal quantum number \(m\), this produces inverse-square singular
terms in the corresponding radial operators.

The curvature invariants in the open cells
\[
I_n=\left(\frac{n\pi}{a},\frac{(n+1)\pi}{a}\right)
\]
are
\[
\mathcal{R}=-4\Lambda,\quad
\mathcal{R}_{\mu\nu}R^{\mu\nu}=8\Lambda^2,\quad
\mathcal{R}_{\mu\nu\alpha\beta}\mathcal{R}^{\mu\nu\alpha\beta}=16\Lambda^2 .
\]
Thus, the singular character of the endpoints is not associated with a
divergence of these scalar curvature invariants inside the cells, but
with the degeneration of the cylindrical angular orbit. Near a zero
\(r=r_n\), one has
\[
S(r)\simeq \sigma a(-1)^n(r-r_n),
\]
so that the spatial two-dimensional \((r,\phi)\) sector has the local
conical form
\[
d\ell^2\simeq dr^2+(\sigma a)^2(r-r_n)^2d\phi^2 .
\]
Unless \(\sigma a=1\), the endpoint may also be associated with a conical
distributional contribution.

Therefore, once the singular endpoints are removed, the radial
manifold is decomposed into open cells \(I_n\). In the present work we
formulate the quantum problem on the first fundamental cell,
\[
0<r<\frac{\pi}{a}.
\]
Equivalently, the same local analysis could be carried out on any
interval between two consecutive zeros of \(S(r)\). The restriction to
one cell is therefore not a merely technical consequence of the
trigonometric form of the radial equations, but follows from the
global structure of the geometry. The other cells are locally
equivalent to the fundamental one, up to a shift in the radial
coordinate and a possible orientation convention for the angular
frame. Hence, a global operator on the punctured half-line is naturally
interpreted as the direct sum of the self-adjoint radial operators
defined on the individual cells. Matching conditions across the
singular endpoints would correspond to an additional physical
assumption, not fixed by the local wave equation, and are not imposed
in the present analysis. This is the geometrical origin of the
singular Sturm--Liouville problems and of the purely discrete radial
spectra obtained below.

With these geometric ingredients at hand, we are now in a position to analyze separately the projected scalar and vector equations in the Bonnor--Melvin--$\Lambda$ spacetime.

\section{Exact scalar sector in the Bonnor--Melvin--$\Lambda$ spacetime}
\label{sec:scalar_sector}

We now apply the projected spin-0 formalism to the Bonnor--Melvin--$\Lambda$
background described in Sect.~\ref{sec:bm_geometry}. The projected scalar field
defined in Eq.~\eqref{eq:phi_definition} satisfies the Klein--Gordon-type equation \eqref{eq:KG_projected}.

For the metric \eqref{eq:BM_metric_compact} the scalar Laplace--Beltrami operator is
\begin{equation}
\nabla_\mu\nabla^\mu
=
-\partial_t^2+\partial_r^2+\frac{S'(r)}{S(r)}\partial_r
+\frac{1}{S^2(r)}\partial_\phi^2+\partial_z^2.
\label{eq:scalar_box}
\end{equation}
Since
\begin{equation}
\frac{S^{\prime}(r)}{S(r)}=a\cot(ar),
\label{eq:Sprime_over_S}
\end{equation}
Eq.~\eqref{eq:KG_projected} becomes
\begin{equation}
\left[
-\partial_t^2+\partial_r^2+a\cot(ar)\,\partial_r
+\frac{1}{\sigma^2\sin^2(ar)}\partial_\phi^2
+\partial_z^2+M^2
\right]\varphi=0.
\label{eq:scalar_wave}
\end{equation}
Using the stationarity, cylindrical symmetry, and translational invariance along
the \(z\)-direction, we adopt the decomposition
\begin{equation}
\varphi(t,r,\phi,z)=e^{-iEt}e^{im\phi}e^{ik_z z}R(r),
\label{eq:scalar_ansatz}
\end{equation}
with $m\in\mathbb{Z}$. Substituting Eq.~\eqref{eq:scalar_ansatz} into Eq.~\eqref{eq:scalar_wave}, the radial function
\(R(r)\) satisfies
\begin{equation}
R''(r)+a\cot(ar)\,R'(r)
+\left[
\kappa^2-\frac{m^2}{\sigma^2\sin^2(ar)}
\right]R(r)=0,
\label{eq:scalar_radial_R}
\end{equation}
where
\begin{equation}
\kappa^2=E^2-M^2-k_z^2.
\label{eq:kappa_scalar}
\end{equation}
As discussed in Sect.~\ref{sec:bm_geometry}, the zeros $r_n=\frac{n\pi}{a}$ of the metric function are singular endpoints of the radial geometry. We therefore solve the scalar radial problem on the fundamental cell
\begin{equation}
0<r<\frac{\pi}{a}.
\label{eq:scalar_fund_interval}
\end{equation}

Moreover, taking the temporal component of the projected scalar current \eqref{eq:projected_scalar_current2} together with Eq.~\eqref{eq:scalar_ansatz}, we obtain the charge density
\begin{equation}
J^0=\frac{E}{M}|R(r)|^2,
\label{eq:scalar_J0}
\end{equation}
since the angular and longitudinal factors have unit modulus. The normalization condition
\begin{equation}
\int d\tau\,J^0=\pm1,
\end{equation}
implies
\begin{equation}
\frac{|E|}{M}\int d\tau\,|R(r)|^2=1.
\end{equation}
Using the spatial volume element induced by the metric $d\tau=S(r)\,dr\,d\phi\,dz$, factoring out the trivial angular dependence and adopting the corresponding longitudinal normalization, one is led to the radial condition
\begin{equation}
\int_0^{\pi/a}S(r)\,|R(r)|^2\,dr<\infty.
\label{eq:scalar_norm_R}
\end{equation}

To cast Eq.~\eqref{eq:scalar_radial_R} into Schr\"odinger-like form, we eliminate the first-derivative term through the transformation
\begin{equation}
R(r)=\frac{u(r)}{\sqrt{\sin(ar)}}.
\label{eq:scalar_R_to_u}
\end{equation}
A direct substitution yields
\begin{equation}
u''(r)+\left[
\kappa^2+\frac{a^2}{4}
-\left(\frac{m^2}{\sigma^2}-\frac{a^2}{4}\right)\csc^2(ar)
\right]u(r)=0.
\label{eq:scalar_u_r}
\end{equation}
Introducing the dimensionless variable and parameters
\begin{align}
x & =ar\in(0,\pi), \label{eq:scalar_x}\\
\lambda &=\nu+\frac12,\label{eq:scalar_lambda}\\
\nu &=\frac{|m|}{\sigma\sqrt{2\Lambda}},\label{eq:scalar_mu}
\end{align}
Eq.~\eqref{eq:scalar_u_r} becomes
\begin{equation}
\left[-\frac{d^2}{dx^2}+\lambda(\lambda-1)\csc^2x\right]u(x)=\varepsilon^{2}u(x),
\label{eq:scalar_PT_standard}
\end{equation}
where
\begin{equation}
\varepsilon^2=\frac{\kappa^{2}}{2\Lambda}+\frac14\,.
\label{eq:scalar_epsilon}
\end{equation}
Hence, the resulting problem is governed by a Schr\"odinger-like equation with a trigonometric P\"oschl--Teller effective potential~\cite{GOLDMAN2006}.

\begin{figure}[ht]
\centering
\includegraphics[width=0.47\textwidth]{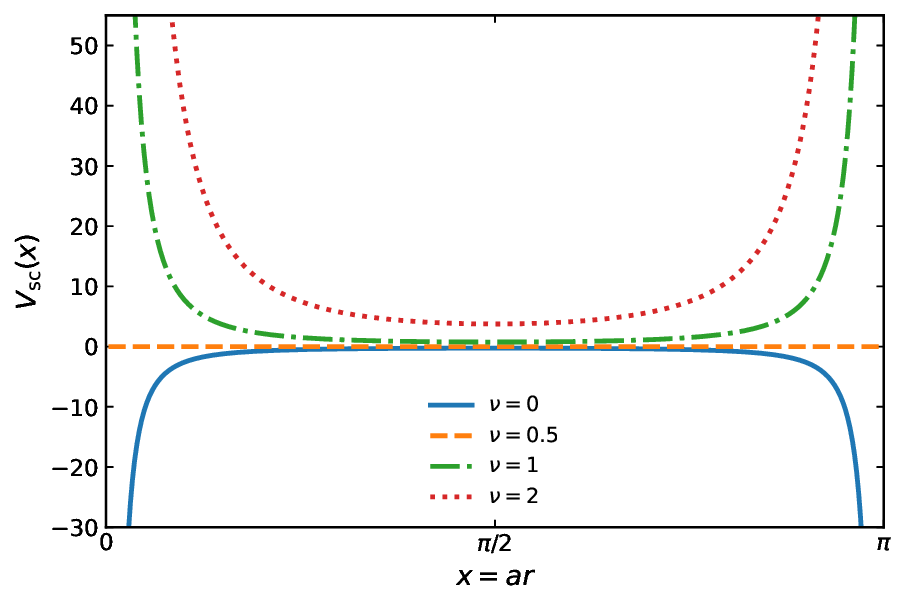}
\caption{Scalar effective potential in the dimensionless variable $x=ar$ for representative values of $\nu$.}
\label{fig:scalar_potential}
\end{figure}

For illustration, Fig.~\ref{fig:scalar_potential} shows the scalar
effective potential in the dimensionless variable \(x=ar\) for representative values of $\nu$. The plot makes explicit how the endpoint behaviour depends on the effective parameter $\nu$. The divergence
at \(x=0\) and \(x=\pi\) reflects the endpoint structure of
the radial Sturm--Liouville problem on the fundamental cell.

The admissibility of the scalar modes is determined by the radial condition \eqref{eq:scalar_norm_R}. Using \eqref{eq:scalar_R_to_u}, this becomes
\begin{equation}
\int_0^\pi |u(x)|^2\,dx<\infty,
\label{eq:scalar_norm_u}
\end{equation}
so that the transformed function belongs to the standard Hilbert space
\(L^2(0,\pi)\). 

At this point, the endpoint conditions must be specified at the level
of the radial operator. The differential expression in Eq.~\eqref{eq:scalar_PT_standard}
defines, on \(C_0^\infty(0,\pi)\), the symmetric operator
\begin{equation}
\mathcal{T}_{\nu}
=
-\frac{d^2}{dx^2}
+
\left(\nu^2-\frac14\right)\csc^2 x\,.
\label{eq:T_nu}
\end{equation}
Since the potential has inverse-square singularities at both endpoints,
the physical problem is not fixed by square integrability alone in all
parameter regimes. We therefore choose the Friedrichs self-adjoint
extension of \(\mathcal{T}_{\nu}\), following the standard characterization of Friedrichs extensions for semibounded singular Sturm--Liouville operators in terms of principal solutions~\cite{PLMSs364:545:1992,AM60:299:2015}. 

Near the singular endpoints, Eq.~\eqref{eq:scalar_PT_standard} is dominated by
the inverse-square term. Thus, near $x=0$,
\begin{eqnarray}
u(x) &\sim & A_0 x^{1/2+\nu}+B_0 x^{1/2-\nu}\,,\label{eq:scalar_local_origin_u}\\
R(x) &\sim & A_0 x^{\nu}+B_0 x^{-\nu}\,,\label{eq:scalar_local_origin_R}
\end{eqnarray}
whereas near $x=\pi$,
\begin{eqnarray}
u(x) &\sim & A_\pi(\pi-x)^{1/2+\nu}+B_\pi(\pi-x)^{1/2-\nu}\,,\label{eq:scalar_local_pi_u}\\
R(x) &\sim & A_\pi(\pi-x)^{\nu}+B_\pi(\pi-x)^{-\nu}\,.\label{eq:scalar_local_pi_R}
\end{eqnarray}
For $0<\nu<1$, both branches of \(u(x)\) are locally square
integrable, so square integrability does not by itself select a unique
self-adjoint realization. The Friedrichs extension selects the
principal branch at each endpoint, namely
\[
B_0=B_\pi=0.
\]
For \(\nu\geq 1\), the more singular branch is not square integrable,
and the same endpoint behaviour is obtained. Therefore, the physical
domain used in this work is the Friedrichs domain, characterized by
\begin{eqnarray}
u(x) &\sim & x^{1/2+\nu}\quad (x\to0),\label{eq:scalar_regular_bc_origin}\\
u(x) &\sim & (\pi-x)^{1/2+\nu}\quad (x\to\pi).\label{eq:scalar_regular_bc_pi}
\end{eqnarray}
In particular, this implies the Dirichlet-type endpoint conditions
\begin{equation}
u(0)=u(\pi)=0.
\label{eq:scalar_dirichlet}
\end{equation}
The critical case \(\nu=0\) requires special attention. In this case,
the two independent local behaviours are $u(x)\sim \sqrt{x}$ and $u(x)\sim \sqrt{x}\ln x$ near $x=0$, and similarly near $x=\pi$. Both branches are square integrable in
\(L^2(0,\pi)\), so normalizability alone is insufficient. The
Friedrichs extension again selects the principal, non-logarithmic
branch and excludes the logarithmically singular solution. Equivalently,
the original radial field $R(x)=u(x)/\sqrt{\sin x}$ remains finite at the endpoints only for the non-logarithmic branch.
Thus, in the critical case the Friedrichs endpoint condition is
\begin{eqnarray}
u(x) & \sim & \sqrt{x}\quad (x\to0), \\
u(x) & \sim & \sqrt{\pi-x}\quad (x\to\pi).
\end{eqnarray}

Using the hypergeometric reduction summarized in \ref{app:hypergeom_spectra_s}, the Friedrichs endpoint conditions lead to the polynomial quantization condition and hence to
\begin{equation}
\varepsilon^{2}_n=(n+\lambda)^2,
\qquad
n=0,1,2,\dots.
\label{eq:scalar_epsilon_spectrum}
\end{equation}
Using Eqs.~\eqref{eq:scalar_epsilon} and \eqref{eq:kappa_scalar}, this yields
\begin{equation}
E_n=\pm\sqrt{M^2+k_z^2+2\Lambda\,(n+\nu)(n+\nu+1)}.
\label{eq:scalar_energy_discussion}
\end{equation}
The energy spectrum exhibits several noteworthy features. First, it is symmetric
under \(E\to -E\), as expected for a relativistic bosonic equation, with the
positive- and negative-energy branches corresponding to particle and
antiparticle solutions, respectively. Second, for fixed \(m\) and \(k_z\), the
radial spectrum is purely discrete, showing that the exact
Bonnor--Melvin--\(\Lambda\) geometry acts as a confining background. This
discretization is therefore induced by the geometry itself, without the need for
an additional external confining interaction.

Another important aspect is that the energy depends only on \(|m|\), so that the
spectrum is invariant under \(m\to -m\). This implies a degeneracy between
states with opposite azimuthal quantum numbers, reflecting the cylindrical
symmetry of the spacetime. On the other hand, the longitudinal momentum \(k_z\)
remains continuous if the \(z\)-direction is not compactified. Hence, the exact
scalar sector exhibits a mixed spectral structure, with discrete radial levels
and a continuous longitudinal degree of freedom.

For the ground state, \(n=0\), one obtains
\begin{equation}
E_0=\pm\sqrt{M^2+k_z^2+2\Lambda\,\nu(\nu+1)}.
\end{equation}
In particular, when \(m=0\), one has \(\nu=0\), and the lowest mode reduces to
\begin{equation}
E_0=\pm\sqrt{M^2+k_z^2}.
\end{equation}
Thus, the ground-state energy receives no additional radial contribution in the
axially symmetric case, whereas for \(m\neq0\) the geometry shifts the spectrum
through the effective angular parameter \(\nu\).

It is also useful to examine the spacing of the levels. For the squared
energies, one finds
\begin{equation}
E_{n+1}^2-E_n^2=4\Lambda(n+\nu+1),
\end{equation}
which shows that the spectrum is not equally spaced. Therefore, unlike the
harmonic oscillator, the separation between successive levels increases with
\(n\). In the large-\(n\) regime (\(n\gg1\)),
\begin{equation}
E_n\sim \pm\sqrt{2\Lambda}\,n,
\end{equation}
so that the asymptotic scale of the radial excitations is set by the
cosmological constant.

The dependence on the geometric parameters becomes more transparent by writing
\begin{equation}
2\Lambda(n+\nu)(n+\nu+1)
=
2\Lambda n(n+1)
+\frac{\sqrt{2\Lambda}\,|m|}{\sigma}(2n+1)
+\frac{m^2}{\sigma^2}.
\end{equation}
This decomposition shows explicitly that the energy receives three distinct
contributions: a purely radial term proportional to \(\Lambda n(n+1)\), a mixed
term coupling the curvature scale to the azimuthal quantum number, and an
effective centrifugal contribution \(m^2/\sigma^2\).

Finally, the limit \(\Lambda\to0\) is subtle. Although the explicit expression
for \(E_n\) suggests that the \(\Lambda\)-dependent radial contributions vanish
in this limit, the corresponding fundamental interval
\begin{equation}
0<r<\frac{\pi}{\sqrt{2\Lambda}}
\end{equation}
simultaneously expands to infinity. Thus, \(\Lambda\) controls not only the coefficients appearing in the energy eigenvalues, but also the size of the radial cell on which the singular Sturm--Liouville problem is defined. Hence, as \(\Lambda\to0\), the boundary-value problem on a finite interval is transformed into a problem on an unbounded radial domain. This is precisely the regime in which the conical approximation becomes applicable, but it should be understood as a different radial problem rather than as a direct limit of the discrete spectrum obtained on a finite cell. For this reason, the confinement mechanism found in the exact geometry is lost only after a simultaneous change of the operator domain, and not merely by setting \(\Lambda=0\) in the discrete spectrum.

The corresponding regular solutions are conveniently written in terms of
Gegenbauer polynomials \cite{ABRAMOWITZ1965} as
\begin{equation}
u_n(x)=\mathcal{N}_n(\sin x)^{\nu+\frac12}
C_n^{\left(\nu+\frac12\right)}(\cos x),
\label{eq:scalar_u_solution}
\end{equation}
and therefore
\begin{equation}
R_n(r)=\mathcal{N}_n[\sin(ar)]^\nu
C_n^{\left(\nu+\frac12\right)}(\cos(ar)),
\label{eq:scalar_R_solution}
\end{equation}
where \(\mathcal{N}_n\) is fixed by the normalization condition
\eqref{eq:scalar_norm_R}.

The eigenfunctions \eqref{eq:scalar_R_solution} also display a number of
characteristic features. Since the prefactor \([\sin(ar)]^\nu\) does not change
sign inside the fundamental interval \((0,\pi/a)\), the internal zeros of
\(R_n(r)\) are entirely determined by the Gegenbauer polynomial. Therefore, the
state labelled by \(n\) possesses exactly \(n\) nodes in the interval
\(0<r<\pi/a\). Moreover, under the reflection \(r\to \pi/a-r\), one has
\(\sin(ar)\to\sin(ar)\) and \(\cos(ar)\to-\cos(ar)\), so that, by using the
parity property \(C_n^{(\lambda)}(-z)=(-1)^nC_n^{(\lambda)}(z)\), it follows that
\begin{equation}
R_n\!\left(\frac{\pi}{a}-r\right)=(-1)^nR_n(r).
\end{equation}
Hence, the eigenfunctions are alternately symmetric and antisymmetric with
respect to the midpoint \(r=\pi/(2a)\). In particular, the ground state
\((n=0)\) is node-free and proportional to \([\sin(ar)]^\nu\), whereas the
excited states acquire an increasing number of oscillations inside the
fundamental interval.

In addition, the parameter \(\nu\) controls the strength of the effective
potential in Eq.~\eqref{eq:scalar_PT_standard} through the relation $\lambda(\lambda-1)=\nu^2-\frac{1}{4}$. Therefore, increasing \(|m|\)
enhances the effective barrier near the endpoints of the fundamental interval.
This is also reflected in the asymptotic behavior of the eigenfunctions,
\begin{equation}
R_n(r)\sim [\sin(ar)]^\nu,
\end{equation}
which shows that larger values of \(|m|\) suppress the scalar field more
strongly near \(r=0\) and \(r=\pi/a\). As a consequence, the corresponding
charge density becomes increasingly depleted close to the boundaries and more
concentrated around the midpoint \(r=\pi/(2a)\).

\begin{figure}[ht]
\begin{subfigure}{0.45\textwidth}
\centering
\includegraphics[width=\textwidth]{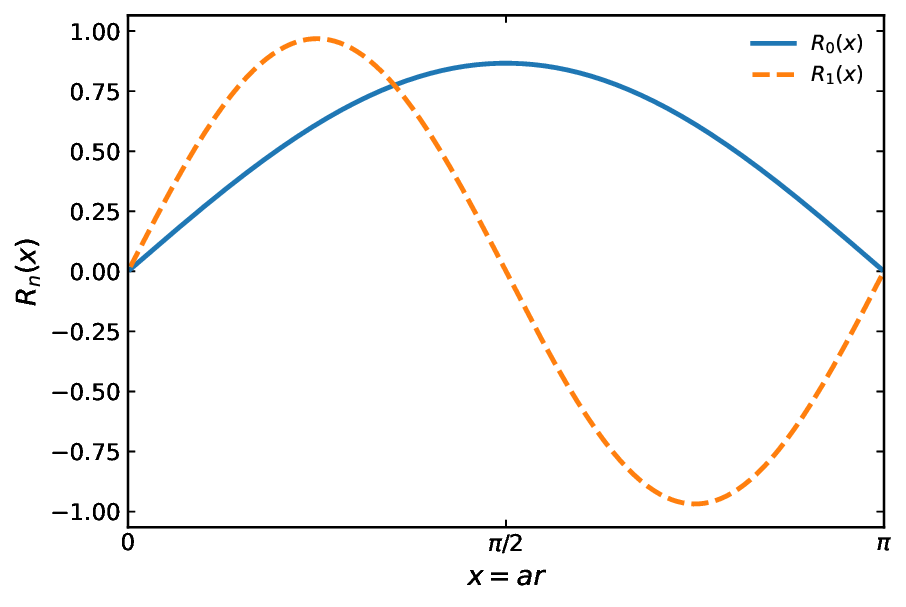}
\caption{}
\end{subfigure}
\begin{subfigure}{0.45\textwidth}
\centering
\includegraphics[width=\textwidth]{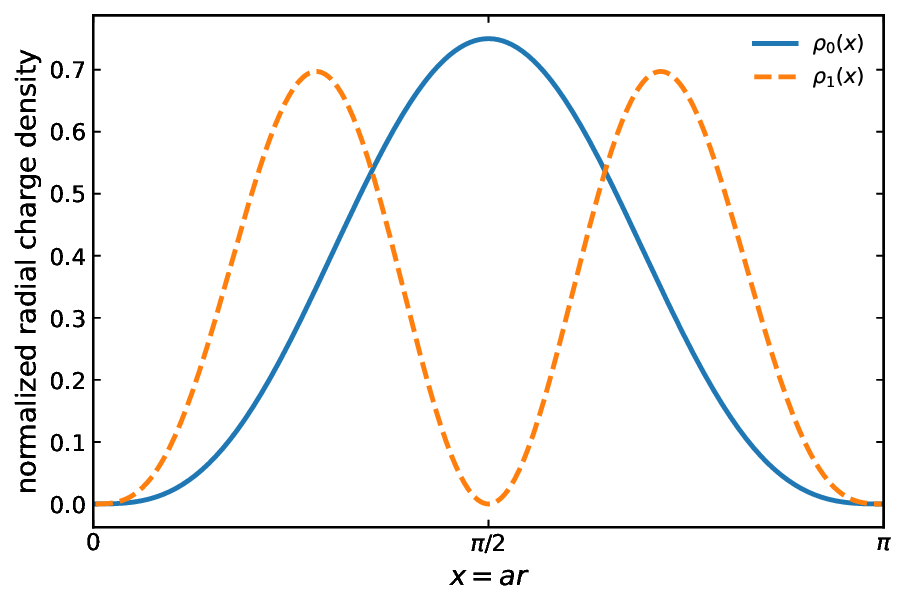}
\caption{}
\end{subfigure}
\caption{Representative scalar radial profiles in the
dimensionless variable \(x=ar\), for \(\nu=1\). (a) Normalized radial
eigenfunctions \(R_n(x)\) for \(n=0,1\). (b) Corresponding normalized
radial charge-density profiles $\rho_{n}(x)$, defined in Eq.~\eqref{eq:scalar_radial_charge_profile}.}
\label{fig:scalar_profiles}
\end{figure}

These features are illustrated in Fig.~\ref{fig:scalar_profiles},
where representative normalized radial eigenfunctions and normalized radial
charge-density profiles are shown for the scalar sector. For a fixed mode, the constant factor $E_{n}/M$ in $J^{0}$ only affects the overall normalization and sign of a given energy branch, but not the shape of the radial distribution. We therefore define the normalized radial charge-density profile as
\begin{equation}
\rho_n(x)=
\frac{\sin x\, |R_n(x)|^2}
{\int_0^\pi \sin x\, |R_n(x)|^2\,dx}.
\label{eq:scalar_radial_charge_profile}
\end{equation}
The plots show the node structure of the
eigenfunctions and the radial confinement inside the fundamental cell.

Therefore, the exact spin-0 sector in the Bonnor--Melvin--$\Lambda$ spacetime is analytically solvable on each fundamental interval between two consecutive zeros of \\ \(\sin(\sqrt{2\Lambda}\,r)\), with exact spectrum and eigenfunctions in closed form. 

\section{Exact vector sector in the Bonnor--Melvin--$\Lambda$ spacetime}
\label{sec:vector_sector}

We now apply the projected spin-1 formalism to the Bonnor--Melvin--$\Lambda$
background described in Sect.~\ref{sec:bm_geometry}. The projected vector field defined in Eq.~\eqref{eq:W_definition}
satisfies the Proca-type equation \eqref{eq:spin1_second_order_reduced},
supplemented by the subsidiary condition \eqref{eq:transversality_condition}.

For convenience, we introduce the orthonormal frame
\begin{equation}
\theta^{\hat 0}=dt,
\quad
\theta^{\hat r}=dr,
\quad
\theta^{\hat \phi}=S(r)\,d\phi,
\quad
\theta^{\hat z}=dz.
\label{eq:orthonormal_frame_append}
\end{equation}
Accordingly, the orthonormal components of the vector field are
\begin{align}\label{eq:orthonormal_components}
W^{\hat t} & =W^t,\\
W^{\hat r} & =W^r,\\
W^{\hat \phi} & =S(r)W^\phi,\\
W^{\hat z} & =W^z,
\end{align}
and introduce the circular combinations
\begin{align}
W_+ & =\frac{1}{\sqrt2}\left(W^{\hat r}+ iW^{\hat \phi}\right),\label{eq:vector_helicity_basis1}\\
W_- & =\frac{1}{\sqrt2}\left(W^{\hat r}- iW^{\hat \phi}\right),\label{eq:vector_helicity_basis2}\\
W_0 & =W^{\hat z},
\end{align}
which correspond to the compact label $s=+1,-1,0$, respectively. The labels \(s=0,\pm1\) have a direct polarization meaning. The mode
\(W_0=W_{\hat z}\) represents the component polarized along the
symmetry axis and will be referred to as the longitudinal mode. By
contrast, \(W_+\) and \(W_-\) are circular combinations of the two
components in the local transverse plane \((\hat r,\hat\phi)\). They
therefore describe opposite transverse circular polarizations, or
equivalently opposite spin weights with respect to local rotations in
this plane.

Strictly speaking, since the radial motion is confined inside a finite
cell, these modes should not be understood as asymptotic plane-wave
helicity eigenstates in the flat-space sense. Nevertheless, because the
background is translationally invariant along the \(z\)-axis, the
decomposition into one longitudinal mode and two transverse circular
polarizations is the natural analogue of the usual helicity
decomposition with respect to the symmetry axis.

In this basis the azimuthal covariant derivative acts diagonally,
\begin{equation}
\nabla_\phi W_s = \left(\partial_\phi + i s\,S'\right)W_s\,.
\label{eq:nabla_phi_ws}
\end{equation}
This diagonal form makes explicit how the spin connection couples to
the circular polarization label \(s\), producing after separation of variables the combination \(m+sS'(r)\) that governs the polarization-dependent transverse effective potentials. 

The vector Laplacian is diagonal in the same polarization basis and is given by
\begin{equation}
\nabla_\mu \nabla^\mu W_s
=
\left[
\partial_t^2
-\partial_r^2
-\frac{S'}{S}\partial_r
-\frac{1}{S^2}\left(\partial_\phi+i s\,S'\right)^2
-\partial_z^2
\right]W_s .
\label{eq:box_ws}
\end{equation}
Moreover, the Ricci term is diagonal in the same basis and may be written as
\begin{equation}
\mathcal{R}_{\hat b}{}^{\hat a}W^{\hat b}
=
s^2\,\frac{S''}{S}\,W_s .
\label{eq:ricci_ws}
\end{equation}
Substituting Eqs.~\eqref{eq:box_ws} and \eqref{eq:ricci_ws} into Eq.~\eqref{eq:spin1_second_order_reduced}, one obtains the compact field equation
\begin{equation}
\begin{split}
\left[
\partial_t^2
-\partial_r^2
-\frac{S'}{S}\partial_r
-\frac{1}{S^2}\left(\partial_\phi+i s\,S'\right)^2\right.\\
\left.-\partial_z^2
+M^2
-s^2\frac{S''}{S}
\right]W_s=0,
\end{split}
\label{eq:field_eq_compact_s}
\end{equation}
with $s=0,\pm1$.

Using stationarity, cylindrical symmetry, and translational invariance along the
\(z\)-direction, we adopt the decomposition
\begin{equation}
W_s(t,r,\phi,z)=e^{-iEt}e^{im\phi}e^{ik_z z}w_{s}(r),
\label{eq:vector_ansatz}
\end{equation}
with \(m\in\mathbb{Z}\). Then Eq.~\eqref{eq:field_eq_compact_s} reduces to
\begin{equation}
\left[
\frac{d^2}{dr^2}
+\frac{S'}{S}\frac{d}{dr}
+\kappa^{2}
-\frac{\left(m+sS'\right)^2}{S^2}
+s^2\frac{S''}{S}
\right]w_s(r)=0.
\label{eq:radial_eq_compact_s}
\end{equation} 

Introducing the dimensionless variable and para\-me\-ter
\begin{align}
x & =ar\in(0,\pi), \label{eq:vector_x}\\
b &=\frac{m}{\sigma\sqrt{2\Lambda}},\label{eq:vector_b}\,
\end{align}
and using \eqref{eq:Sprime_over_S}, together with
\begin{equation}
\frac{S''}{S}=-a^2,
\quad
\frac{m+sS'}{S}=a\,\frac{b+s\cos x}{\sin x},
\label{eq:useful_S_relations}
\end{equation}
Eq.~\eqref{eq:radial_eq_compact_s} becomes
\begin{equation}
\left[
\frac{d^2}{dx^2}
+\cot x\,\frac{d}{dx}
+\frac{\kappa^{2}}{a^{2}}
-\frac{(b+s\cos x)^2}{\sin^2 x}
-s^2
\right]w_s(x)=0 .
\label{eq:radial_x_compact}
\end{equation}

The charge density for the vector sector follows from the temporal part of the projected current \eqref{eq:projected_vector_current}, which yields
\begin{equation}
\begin{split}
J^{0} &= \frac{E}{M}\left( \vert w^{\hat{r}}\vert^{2}+\vert w^{\hat{\phi}}\vert^{2}+\vert w^{\hat{z}}\vert^{2} \right)
-\frac{1}{M}\Im\left[ (w^{\hat{r}})^{\ast}\partial_{r}w^{\hat{t}} \right]\\
&-\frac{m}{MS}\Re\left[ (w^{\hat{\phi}})^{\ast}w^{\hat{t}} \right]-\frac{k_{z}}{M}\Re\left[ (w^{\hat{z}})^{\ast}w^{\hat{t}} \right],
\end{split}
\label{eq:J0_geral}
\end{equation}
as discussed in \ref{app:spin1_normalization}.

To cast Eq.~\eqref{eq:radial_x_compact} into Schr\"odinger-like form, we write
\begin{equation}
w_s(x)=\frac{u_s(x)}{\sqrt{\sin x}}.
\label{eq:w_to_u}
\end{equation}
This yields
\begin{equation}
\left[
-\frac{d^2}{dx^2}
+\frac{(b+s\cos x)^2-\frac14}{\sin^2 x}+s^{2}
\right]u_s(x)
=
\varepsilon^{2}u_s(x),
\label{eq:Schrodinger_Scarf}
\end{equation}
where $\varepsilon^{2}$ is given by Eq.~\eqref{eq:scalar_epsilon}.
Using the identities 
\begin{equation}
\sin^{2}x=4\sin^{2}\frac{x}{2}\cos^{2}\frac{x}{2}, \quad \cos x=\cos^{2}\frac{x}{2}-\sin^{2}\frac{x}{2},
\label{eq:PT_identity}
\end{equation}
Eq.~\eqref{eq:Schrodinger_Scarf} can be rewritten as
\begin{equation}
\left[
-\frac{d^2}{dx^2}
+\frac{\alpha_{s}^{2}-\frac14}{4\sin^2(x/2)}
+\frac{\beta_{s}^{2}-\frac14}{4\cos^2(x/2)}
\right]u_s(x)
=\varepsilon^{2}u_s(x),
\label{eq:PT_compact_s}
\end{equation} 
where $\alpha_{s}=\vert b+s\vert$ and $\beta_{s}=\vert b-s\vert$. Hence, the vector sector is equivalently described by a Sch\"odinger-like equation with generalized trigonometric P\"oschl--Teller potential \cite{ZP83:143:1933,FLUGGE1999}. Inherently linked to the vectorial nature of the DKP field, the system's degrees of freedom naturally decouple into distinct physical sectors. In the following subsections, we perform a detailed analysis of these solutions by addressing the longitudinal and transverse modes separately. Specifically, we first examine the case $s=0$, representing the longitudinal component, followed by the $s=\pm 1$ cases, which describe the transverse dynamics of the field.

\subsection{Longitudinal mode}

For the longitudinal sector ($s=0$), we have $\alpha_{0}=\beta_{0}=\vert b\vert=\nu$ and using the identities \eqref{eq:PT_identity}, Eq.~\eqref{eq:PT_compact_s} reduces to
\begin{equation}
\left[-\frac{d^2}{dx^2}+\lambda(\lambda-1)\csc^2x\right]u_{0}(x)=\varepsilon^{2}u_{0}(x).
\label{eq:PT_longitudinal}
\end{equation}
This is exactly the same effective equation obtained in the scalar sector,
Eq.~\eqref{eq:scalar_PT_standard}. The reason is that the longitudinal
component \(W_0=W^{\hat z}\) carries no transverse spin weight with
respect to local rotations in the \((\hat r,\hat\phi)\) plane.
Consequently, the spin-connection shift \(m+sS'(r)\) reduces simply to
\(m\), and the Ricci contribution proportional to \(s^2\) is absent.
The longitudinal polarization therefore probes the same radial
Sturm--Liouville operator as the scalar sector. The longitudinal spin-1 mode is thus
governed by the same trigonometric P\"oschl--Teller problem, with identical
admissibility conditions, spectrum, and eigenfunctions after the replacement
\(u(x)\to u_{0}(x)\) \(\left[R(r)\to w_{0}(r)\right]\). In particular, the admissibility of the longitudinal mode is determined by the radial condition
\begin{equation}
\int_0^{\pi/a}S(r)\vert w_{0}(r)\vert^{2}dr<\infty,
\label{eq:vector_longitudinal_norm}
\end{equation}
whose derivation is presented in \ref{app:longitudinal_normalization}. 

Hence, the longitudinal mode is exactly isospectral with the scalar sector, and its energy spectrum is
\begin{equation}
E_{0,n}=\pm\sqrt{M^2+k_z^2+2\Lambda\,(n+\nu)(n+\nu+1)},
\label{eq:vector_longitudinal_energy}
\end{equation}
with $n=0,1,2,\dots$, and the corresponding exact solutions are
\begin{equation}
w_{0,n}(r)=\mathcal{N}_{0,n}[\sin(ar)]^\nu
C_n^{\left(\nu+\frac12\right)}(\cos(ar)),
\label{eq:vector_longitudinal_F}
\end{equation}
where $\mathcal{N}_{0,n}$ is fixed by the normalization condition \eqref{eq:vector_longitudinal_norm}.

Since the longitudinal eigenfunctions are obtained from the scalar
ones by the replacement \(R(r)\to w_0(r)\), the radial profiles shown
in Fig.~\ref{fig:scalar_profiles} also illustrate the behaviour
of the longitudinal vector mode. Likewise, the effective potential governing the longitudinal mode is exactly the same as the scalar one shown in Fig.~\ref{fig:scalar_potential}.

\subsection{Transverse modes}

In the transverse sector, the functions \(w_s(r)\) (with \(s=\pm1\)) satisfy Eq.~\eqref{eq:PT_compact_s}. Consequently, these modes are governed by a Schr\"odinger-like equation featuring a generalized trigonometric P\"oschl--Teller potential \cite{ZP83:143:1933,FLUGGE1999}. 

\begin{figure}[ht]
\begin{subfigure}{0.45\textwidth}
\centering
\includegraphics[width=\textwidth]{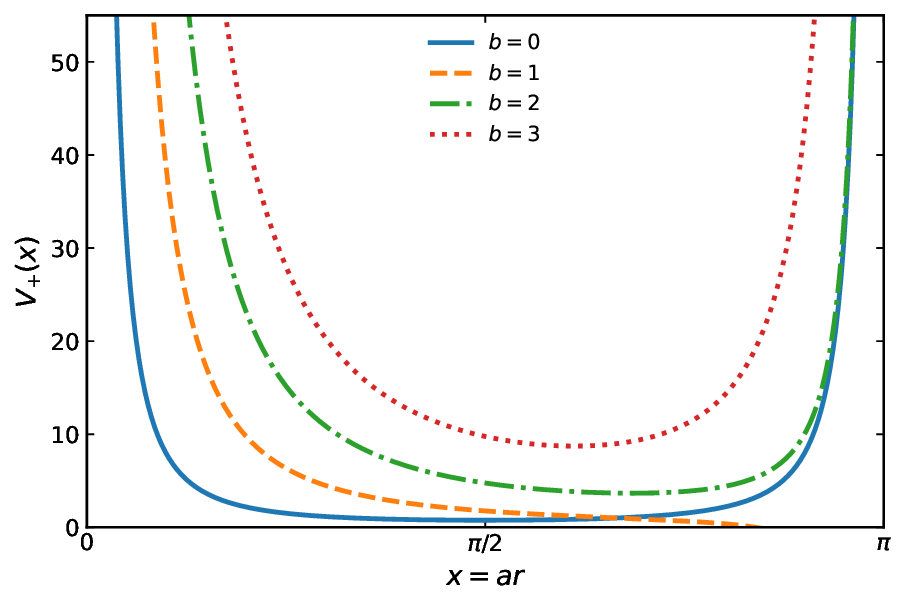}
\caption{}
\end{subfigure}
\begin{subfigure}{0.45\textwidth}
\centering
\includegraphics[width=\textwidth]{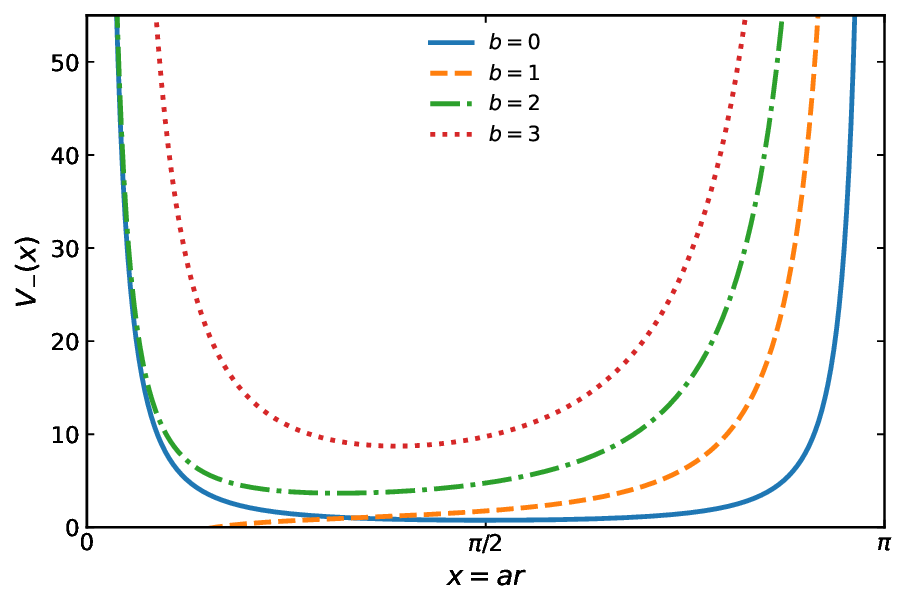}
\caption{}
\end{subfigure}
\caption{Effective potentials for the transverse circular polarizations in the
dimensionless variable \(x=ar\), for representative values of \(b\). (a) $s=+1$. (b) $s=-1$.}
\label{fig:transverse_potential}
\end{figure}

For illustration, Fig.~\ref{fig:transverse_potential} shows the
effective potentials associated with the transverse circular
polarizations in the dimensionless variable \(x=ar\), for
representative values of \(b\). Panel (a) corresponds to \(s=+1\),
whereas panel (b) corresponds to \(s=-1\). The plots make explicit the dependence of the endpoint behaviour on the effective parameters
\(\alpha_s=|b+s|\) and \(\beta_s=|b-s|\). 

Reducing the projected current within the same circular polarization sector reveals that, once the trivial angular and longitudinal dependencies are factored out, the admissibility of the transverse modes is governed by
\begin{equation}
\int_0^{\pi/a}S(r)\vert w_s(r)\vert^{2}dr<\infty,
\qquad s=\pm1,
\label{eq:vector_transverse_norm}
\end{equation}
as discussed in \ref{app:transverse_normalization}. Using \eqref{eq:w_to_u}, Eq.~\eqref{eq:vector_transverse_norm} can be written equivalently as
\begin{equation}
\int_0^{\pi}\vert u_{s}(x)\vert^{2}dx<\infty.
\label{eq:vector_transverse_norm_u}
\end{equation}

The corresponding transverse differential expression in Eq.~\eqref{eq:PT_compact_s} defines, on
\(C_0^\infty(0,\pi)\), the symmetric operator
\begin{equation}
\mathcal{T}_{\alpha_s,\beta_s}
=
-\frac{d^2}{dx^2}
+
\frac{\alpha_s^2-\frac14}{4\sin^2(x/2)}
+
\frac{\beta_s^2-\frac14}{4\cos^2(x/2)}\,.
\label{eq:T_alpha_beta}
\end{equation}
As in the scalar sector, the physical radial realization is chosen as the Friedrichs self-adjoint
extension of $\mathcal{T}_{\alpha_s,\beta_s}$. The corresponding endpoint behavior follows from the leading inverse-square terms. Near
\(x=0\), Eq.~\eqref{eq:PT_compact_s} implies
\begin{equation}
u_s(x)\sim x^{\frac12\pm\alpha_s},
\qquad
x\to0,
\label{eq:vector_transverse_local_0}
\end{equation}
whereas near \(x=\pi\),
\begin{equation}
u_s(x)\sim (\pi-x)^{\frac12\pm\beta_s},
\qquad
x\to\pi.
\label{eq:vector_transverse_local_pi}
\end{equation}
Therefore, the Friedrichs prescription selects the principal endpoint branches
\begin{align}\label{eq:vector_transverse_regular_bc}
u_s(x) &\propto x^{\frac12+\alpha_s} \qquad (x\to0),\\
u_s(x) &\propto (\pi-x)^{\frac12+\beta_s} \qquad (x\to\pi)\,,
\end{align}
and excludes the logarithmic branches in the critical cases $\alpha_{s}=0$ or $\beta_{s}=0$.

Using the analogous hypergeometric reduction summarized in~\ref{app:hypergeom_spectra_v}, the Friedrichs realization yields the polynomial quantization condition
\begin{equation}
\varepsilon_{s,n}^{2}
=
\left(
n+\frac{\alpha_s+\beta_s+1}{2}
\right)^2,
\qquad
n=0,1,2,\dots,
\label{eq:vector_transverse_spectrum}
\end{equation}
with $s=\pm1$. Using Eq.~\eqref{eq:scalar_epsilon}, this yields
\begin{equation}
E_{s,n}
=
\pm\sqrt{
M^2+k_z^2
+
2\Lambda\left[
\left(
n+\frac{\alpha_s+\beta_s+1}{2}
\right)^2-\frac14
\right]
}.
\label{eq:vector_transverse_energy}
\end{equation}
The transverse energy spectrum also exhibits a number of characteristic
features. First, it is symmetric under \(E\to -E\), as expected for a
relativistic bosonic equation, with the positive- and negative-energy branches
corresponding to particle and antiparticle solutions, respectively. Second, in
contrast with the longitudinal sector, the transverse spectrum is controlled by
the effective parameters $\alpha_s$, $\beta_s$ and $b$, which encode the coupling between polarization and geometry.

A noteworthy point is that, although the effective potentials for the
two circular polarizations are different, their energy eigenvalues
remain degenerate. Indeed, under \(s\to -s\), the parameters
\(\alpha_s\) and \(\beta_s\) are interchanged, so that the combination
\(\alpha_s+\beta_s\) entering the spectrum is unchanged. Therefore,
\begin{equation}
E_{+,n}=E_{-,n},
\end{equation}
and the two transverse circular polarizations are not split
energetically. 

Moreover, since
\begin{equation}
\alpha_s+\beta_s=2\max(|b|,1),
\end{equation}
the transverse spectrum may be written in the compact form
\begin{equation}
E_{s,n}
=
\pm\sqrt{
M^2+k_z^2
+
2\Lambda\left[
\left(
n+\max(|b|,1)+\frac12
\right)^2-\frac14
\right]
}.
\end{equation}
This expression shows that the transverse sector has two distinct regimes. For
\(|b|\leq 1\), the spectrum becomes independent of \(m\),
\begin{equation}
E_{s,n}
=
\pm\sqrt{M^2+k_z^2+2\Lambda\,(n+1)(n+2)},
\end{equation}
whereas for \(|b|\geq 1\) one finds
\begin{equation}
E_{s,n}
=
\pm\sqrt{M^2+k_z^2+2\Lambda\,(n+|b|)(n+|b|+1)}.
\end{equation}
Thus, beyond the threshold \(|m|=\sigma\sqrt{2\Lambda}\), the transverse energy
levels recover a similar functional dependence on the effective angular
parameter as in the scalar and longitudinal sectors.

The corresponding regular solutions can be written in terms of Jacobi
polynomials as
\begin{equation}
u_{s,n}(x)
=
\mathcal{N}_{s,n}
\left(\sin\frac{x}{2}\right)^{\alpha_s+\frac12}
\left(\cos\frac{x}{2}\right)^{\beta_s+\frac12}
P_n^{(\alpha_s,\beta_s)}(\cos x),
\label{eq:vector_transverse_u}
\end{equation}
and therefore
\begin{equation}
w_{s,n}(r)
=
\widetilde{\mathcal{N}}_{s,n}
\left(\sin\frac{ar}{2}\right)^{\alpha_s}
\left(\cos\frac{ar}{2}\right)^{\beta_s}
P_n^{(\alpha_s,\beta_s)}(\cos(ar)),
\label{eq:vector_transverse_F}
\end{equation}
with $s=\pm1$ and where the normalization constants are fixed by
Eq.~\eqref{eq:vector_transverse_norm}. The transverse eigenfunctions also display a rich structure. Since the Jacobi
polynomial \(P_n^{(\alpha_s,\beta_s)}(z)\) has exactly \(n\) simple zeros in the
interval \(z\in(-1,1)\), the state labelled by \(n\) possesses exactly \(n\)
internal nodes in the fundamental interval \(0<r<\pi/a\). Therefore, the radial
quantum number retains the same Sturm--Liouville interpretation as in the
scalar and longitudinal cases.

The endpoint behavior is controlled by the exponents \(\alpha_s\) and
\(\beta_s\). Indeed, near \(r=0\),
\begin{equation}
w_{s,n}(r)\sim \left(\sin\frac{ar}{2}\right)^{\alpha_s},
\end{equation}
whereas near \(r=\pi/a\),
\begin{equation}
w_{s,n}(r)\sim \left(\cos\frac{ar}{2}\right)^{\beta_s}.
\end{equation}
Hence, the two transverse polarizations are generally suppressed differently at
the two endpoints of the interval. When \(\alpha_s\neq\beta_s\), the
corresponding radial charge density becomes asymmetrically distributed inside the
fundamental cell, being pushed away more strongly from one boundary than from
the other.

This asymmetry is reversed under \(s\to -s\). In fact, since
\begin{equation}
\alpha_{-s}=\beta_s,
\qquad
\beta_{-s}=\alpha_s,
\end{equation}
and the Jacobi polynomials satisfy
\begin{equation}
P_n^{(\alpha,\beta)}(-z)=(-1)^nP_n^{(\beta,\alpha)}(z),
\end{equation}
one finds that the two transverse circular polarizations are related by reflection about
the midpoint of the fundamental interval,
\begin{equation}
w_{s,n}\!\left(\frac{\pi}{a}-r\right)\propto (-1)^{n}w_{-s,n}(r),
\end{equation}
up to the choice of normalization constants. Therefore, the two circular
polarizations have the same energy spectrum, but their radial profiles are
mirror images of one another.

\begin{figure}[ht]
\begin{subfigure}{0.45\textwidth}
\centering
\includegraphics[width=\textwidth]{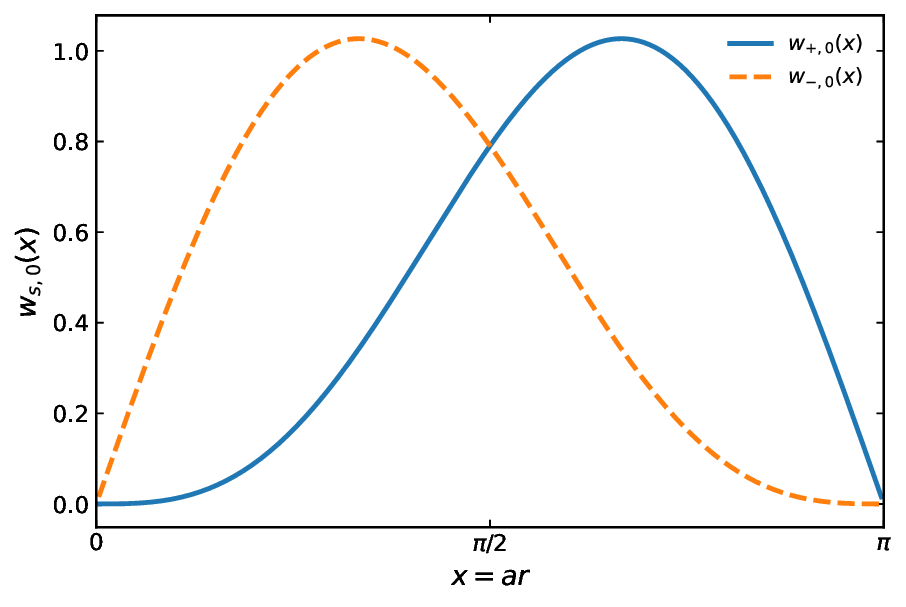}
\caption{}
\end{subfigure}
\begin{subfigure}{0.45\textwidth}
\centering
\includegraphics[width=\textwidth]{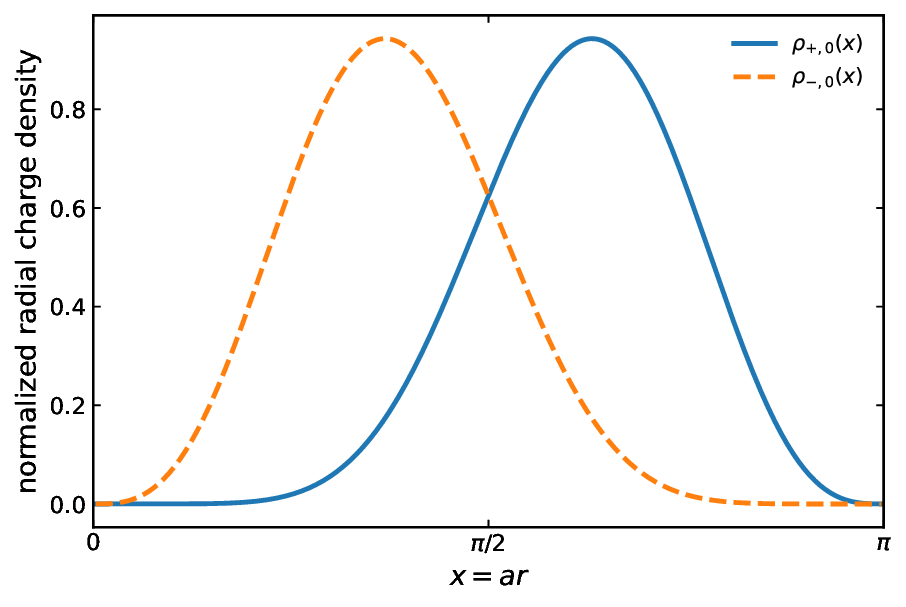}
\caption{}
\end{subfigure}
\caption{Representative transverse radial profiles in the
dimensionless variable \(x=ar\), for \(b=2\) and $n=0$. (a) Normalized radial
eigenfunctions \(w_{+,0}(x)\) and \(w_{-,0}(x)\). (b) Corresponding normalized
radial charge-density profiles $\rho_{+,0}(x)$ and $\rho_{-,0}(x)$, defined in Eq.~\eqref{eq:transverse_radial_charge_profile}.}
\label{fig:transverse_profiles}
\end{figure}

These features are illustrated in Fig.~\ref{fig:transverse_profiles},
where representative normalized radial eigenfunctions and normalized
radial charge-density profiles are shown for the transverse sector.
For a fixed circular-polarization mode, the projected charge density \(J^0_{(s)}\), derived in \ref{app:transverse_normalization}, is proportional to \(|w_{s,n}(r)|^2\) up to a constant prefactor. This prefactor 
only affects the overall normalization and, for the negative-energy branch, the overall sign of \(J^0_{(s)}\), but not the shape of the radial distribution. We therefore
define the normalized transverse radial charge-density profile as
\begin{equation}
\rho_{s,n}(x)=
\frac{\sin x\, |w_{s,n}(x)|^2}
{\int_0^\pi \sin x\, |w_{s,n}(x)|^2\,dx},
\label{eq:transverse_radial_charge_profile}
\end{equation}
with $s=\pm1$. The plots show that the two transverse circular polarizations are related by reflection about the midpoint \(x=\pi/2\), illustrating the polarization-dependent radial localization inside the fundamental cell, even though
their energy eigenvalues are degenerate.

In the special case \(m=0\), one has \(b=0\), so that
\begin{equation}
\alpha_s=\beta_s=1,
\qquad s=\pm1.
\end{equation}
Hence, both transverse polarizations are described by the same symmetric Jacobi
family, and the endpoint suppression is identical at the two boundaries. For
\(m\neq0\), however, the difference between \(\alpha_s\) and \(\beta_s\)
introduces an intrinsic left-right asymmetry in the transverse radial profile,
even though the corresponding energy eigenvalues remain unchanged.

Unlike the longitudinal mode, the transverse sector is generally not
isospectral with the scalar sector. The reason is that the label
\(s=\pm1\) carries the transverse circular-polarization information,
whereas \(m\) characterizes the orbital dependence \(e^{im\phi}\).
The spin connection combines these two structures through the
polarization-dependent term \(m+sS'(r)\), or equivalently through the
endpoint parameters \(|b+s|\) and \(|b-s|\). In addition, the Ricci
tensor contributes explicitly to the transverse radial equation. As a
consequence, the transverse sector belongs to a distinct solvable
family of effective potentials. Although the two transverse circular
polarizations remain degenerate at the level of the energy spectrum,
their radial profiles are different and are related by reflection
within the fundamental interval.

The radial profiles discussed above also clarify the spatial
interpretation of the bound states. Since the radial coordinate is
restricted to a fundamental cell, whereas the azimuthal coordinate
remains periodic, the modes obtained here are radially
confined but not angularly localized. For separated solutions of the
form \(e^{-iEt}e^{im\phi}e^{ik_z z}\) times a radial amplitude, the local charge density is
independent of $t$, $\phi$, and $z$, since the temporal, azimuthal, and longitudinal factors have unit modulus. After factoring out the trivial angular and longitudinal dependence, it therefore depends only on the radial coordinate inside the fundamental cell. Thus, at fixed
\(z\), the charge distribution extends along the whole angular orbit
and forms a ring-like profile in the transverse section.

Because the solutions are also extended along the symmetry axis through
the factor \(e^{ik_z z}\), their global spatial structure is more
accurately described as cylindrical or tube-like rather than as a
localized three-dimen\-sional ring. In this sense, the case \(m=0\)
corresponds to a stationary ring-like configuration with no orbital
circulation around the axis, whereas states with \(m\neq0\) represent
stationary circulating ring-like modes with quantized azimuthal
dependence.

Therefore, the exact spin-1 sector in the Bonnor--Melvin--$\Lambda$ spacetime
remains analytically tractable. The longitudinal mode is governed by the same
trigonometric P\"oschl--Teller potential as the scalar sector, whereas the
transverse circular polarizations are described by generalized trigonometric P\"oschl--Teller potentials. In all cases, the physically admissible solutions are selected by the Friedrichs self-adjoint realization of the corresponding singular
radial operators on the fundamental interval \(0<r<\pi/a\).

\section{Conclusions}
\label{sec:conclusions}

In this work, we investigated scalar and vector bosons in the
Bonnor--Melvin--$\Lambda$ spacetime within the Duffin--Kemmer--Petiau (DKP) formalism.
Using Ume\-zawa's projection operators, we isolated the physical spin-0
and spin-1 sectors and derived the corresponding exact second-order equations
in the full curved spacetime, without resorting to the conical
approximation. Consequently, the projected DKP approach reproduces the expected Klein--Gordon and Proca-type equations, together with their
associated conserved currents, within a unified framework.

For the scalar sector, we showed that the radial equation can be mapped onto a Schr\"odinger-like problem with a trigonometric P\"oschl--Teller effective potential. The zeros of the metric function divide the radial domain into fundamental intervals, so that the system is naturally formulated as a singular Sturm--Liouville problem on a finite interval. The physical radial domain is fixed by the Friedrichs self-adjoint
extension of the corresponding singular Sturm--Liouville operator,
which selects the principal endpoint behaviour and yields a purely
discrete radial spectrum, with exact eigenfunctions expressed in terms of Gegenbauer polynomials.

For the vector sector, we obtained an equally consistent exact treatment. The longitudinal polarization is governed by the same trigonometric P\"oschl--Teller potential found in the scalar sector, whereas the transverse
polarizations are described by generalized trigonometric P\"oschl--Teller potentials. Hence, all physical
spin-1 modes remain analytically tractable in the full Bonnor--Melvin--$\Lambda$ geometry, with exact solutions written in terms of Gegenbauer and Jacobi polynomials for the longitudinal and transverse sectors, respectively. In addition, the scalar sector and the longitudinal vector mode are isospectral, while the transverse modes belong to a distinct solvable family. Although the two transverse circular polarizations remain degenerate in energy, their radial profiles differ and are related by reflection within the fundamental interval. The corresponding energy levels depend explicitly on the cosmological constant $\Lambda$ and on the geometric parameter $\sigma$, thereby highlighting the role of the background geometry in shaping the spectral properties of the bosonic states.

A central outcome of the present analysis is that the Bonnor--Melvin--$\Lambda$ spacetime acts as a purely geometric confining
background for both scalar and vector bosons. Unlike the conical approximation, which may lead to an effective problem on the half-line, the exact geometry yields singular boundary-value problems on a fundamental radial cell and, consequently, purely discrete radial spectra. In this sense, confinement is not introduced by an external interaction but is encoded in the global structure of the spacetime itself.

Overall, the present results extend previous approximate treatments and provide a unified exact description of spin-0 and spin-1
bosons in the Bonnor--Melvin--$\Lambda$ spacetime. Beyond their mathematical interest, these findings highlight how curvature and global geometry shape the spectral behavior of relativistic bosonic fields in nontrivial Einstein--Maxwell backgrounds. Natural extensions of the present analysis include the incorporation of additional electromagnetic interactions, the study of observables derived from the exact spectra, and the application of the projected DKP framework to related curved spacetimes.

\section*{Statements and Declarations}

\subsection*{Competing interests}
The authors declare no competing interests.

\subsection*{Data availability}
Data sharing is not applicable to this article as no datasets were generated or analyzed during the current study.

\begin{acknowledgements}
The authors thank the anonymous reviewers for their valuable comments and suggestions, which helped improve the quality and clarity of this paper. This work was supported in part by means of funds provided by CNPq, Brazil, Grant No. 308172/2023-0, FAPEMA and CAPES - Finance code 001.
\end{acknowledgements}

\appendix

\section{Spin-0 subalgebra: $P$-algebra}
\label{A:1}
In this Appendix, we summarize the algebraic properties of the spin-0 subalgebra. Following the notation of Ref.~\cite{JMP14:1760:1973}, the $P$-algebra is defined by the set of operators $\left\{ P,\,^{\mu }\!P,P^{\mu
},\,^{\mu }\!P^{\nu }\right\} $, which satisfy the followings relations:%
\begin{eqnarray}
P\left( P^{\mu }\right)  &=&P^{\mu },\quad \left( ^{\mu }\!P\right) P=\,^{\mu }\!P,
\\
\left( P^{\mu }\right) P &=&P\left( ^{\mu }\!P\right) =0,
\label{eq:P_algebra_main}
\\
\left( ^{\mu }\!P\right) \left( P^{\nu }\right)  &=&\,^{\mu }\!P^{\nu },
\quad
\left( P^{\mu }\right) \left( ^{\nu }\!P\right) =g^{\mu \nu }P.
\end{eqnarray}
Hence%
\begin{eqnarray}
P\left( \,^{\mu }\!P^{\nu }\right) &=&\left( \,^{\mu }\!P^{\nu }\right) P=0,
\\
\left( P^{\mu }\right) \left( P^{\nu }\right)
&=&\left( ^{\nu }\!P\right) \left( ^{\mu }\!P\right) =0,
\\
\beta ^{\mu } &=&P^{\mu }+\,^{\mu }\!P,
\qquad
\bar{\Psi}P=\left( P\Psi \right) ^{\dagger }.
\label{eq:P_algebra_beta}
\end{eqnarray}

\section{Spin-1 subalgebra}
\label{A:2}

For completeness, we provide a summary of the spin-1 subalgebra following the notation of Ref.~\cite{JMP14:1760:1973}. Note that, consistent with the original reference, explicit summations are retained here instead of the standard Einstein convention. The spin-1 sector is characterized by the operators
$\{^{\mu }\!\,V^{\nu },\,^{\mu }\!\,V^{\nu \lambda },\,^{\nu
\lambda }\!\,V^{\mu },\,^{\nu \lambda }\!\,V^{\mu \sigma }\}$
with%
\begin{eqnarray}
^{\mu }\!\,V^{\nu } &=&\left( ^{\mu }\!R\right) \left( \!R^{\nu }\right)
,\quad ^{\mu }\!\,V^{\nu \lambda }=\left( ^{\mu }\!R\right) \left( \!R^{\nu
\lambda }\right),    \\
^{\nu \lambda }\!\,V^{\mu } &=&\left( ^{\nu \lambda }\!R\right) \left(
\!R^{\mu }\right) ,\quad ^{\nu \lambda }\!\,V^{\mu \sigma }=\left( ^{\nu
\lambda }\!R\right) \left( \!R^{\mu \sigma }\right), 
\end{eqnarray}%
where%
\begin{eqnarray}
\left( \!R^{\mu }\right) \left( ^{\nu }\!R\right)  &=&\left( \!R^{0}\right)
g^{\mu \nu },\quad \left( \!R^{\mu }\right) \left( \!R^{\nu \lambda }\right)
=\left( \!R^{\nu \lambda }\right) g^{\mu 0}, \\
\left( \!R^{\mu }\right)
\left( \!R^{\nu }\right) &=&  \left( \!R^{\nu }\right) g^{\mu 0}, \\
\left( \!R^{\mu \nu }\right) \left( ^{\lambda }\!R\right)  &=&\left(
\!R^{\mu }\right) \left( ^{\nu \lambda }\!R\right) =\left( \!R^{\mu \nu
}\right) \left( \!R^{\lambda }\right) =0, \label{v2} \\
\left( \!R^{\mu \nu }\right) \left( ^{\lambda \sigma }\!R\right)  &=&\left(
\!R^{0}\right) \Delta ^{\mu \nu \lambda \sigma },\\
\Delta ^{\mu \nu
\lambda \sigma } &=& g^{\mu \sigma }g^{\nu \lambda }-g^{\mu \lambda }g^{\nu
\sigma }.
\end{eqnarray}%

Using Eq.~\eqref{v2}, one finds%
\begin{eqnarray}
\left( ^{\mu }\!\,V^{\nu \lambda }\right) \left( ^{\rho \sigma }\!\,V^{\tau
}\right)  &=&\left( ^{\mu }\!\,V^{\tau }\right) \Delta ^{\nu \lambda \rho
\sigma },   \\
\left( ^{\mu }\!\,V^{\nu \lambda }\right) \left( ^{\tau }\!\,V^{\rho \sigma
}\right)  &=&\left( ^{\nu \lambda }\!\,V^{\mu }\right) \left( ^{\rho \sigma
}\!\,V^{\tau }\right) =0, \\
\beta ^{\mu } &=&\sum_{\lambda }\left( ^{\mu \lambda }\!\,V^{\lambda
}+\,^{\lambda }\!\,V^{\lambda \mu }\right), \\
\bar{\Psi}\left( \!R^{0}\right)  &=&\left( \!R^{0}\Psi \right) ^{\dagger
}\eta ^{0},\\
\bar{\Psi}\!\left( ^{i0}\!R\right) &=& \left( \!R^{i0}\Psi
\right) ^{\dagger }\eta ^{0}.
\end{eqnarray}%

\section{Hypergeometric derivation of the radial spectra}
\label{app:hypergeom_spectra}

\subsection{Scalar and longitudinal sectors}
\label{app:hypergeom_spectra_s}

The scalar sector and the longitudinal vector mode lead to the
trigonometric Pöschl--Teller equation
\begin{equation}
\left[
-\frac{d^2}{dx^2}
+
\lambda(\lambda-1)\csc^2x
\right]u(x)
=
\varepsilon^2u(x),
\qquad 0<x<\pi ,
\label{app:eq_PT_scalar}
\end{equation}
where \(\lambda=\nu+1/2\). In order to reduce this
equation to hypergeometric form, we introduce
\begin{equation}
z=\frac{1-\cos x}{2},
\qquad 0<z<1,
\label{app:eq_z_scalar}
\end{equation}
and factor out the Friedrichs endpoint behaviour by
writing
\begin{equation}
u(x)=(\sin x)^\lambda F(z).
\label{app:eq_ansatz_scalar}
\end{equation}
Substitution of Eq.~\eqref{app:eq_ansatz_scalar} into
Eq.~\eqref{app:eq_PT_scalar} gives
\begin{equation}
z(1-z)\frac{d^2F}{dz^2}
+
\left[
\lambda+\frac12-(2\lambda+1)z
\right]\frac{dF}{dz}
+
(\varepsilon^2-\lambda^2)F=0 .
\label{app:eq_hyper_scalar}
\end{equation}
This is the Gauss hypergeometric equation
\begin{equation}
z(1-z)F''+[c-(a+b+1)z]F'-abF=0,
\end{equation}
with
\begin{equation}
c=\lambda+\frac12,
\qquad
a=\lambda-\varepsilon,
\qquad
b=\lambda+\varepsilon .
\label{app:eq_ab_scalar}
\end{equation}
The solution regular at \(z=0\) is
\begin{equation}
F(z)={}_2F_1(a,b;c;z).
\end{equation}
The Friedrichs endpoint condition at $z=1$ is satisfied by the polynomial solutions, obtained when the hypergeometric series terminates. This is
achieved by imposing
\begin{equation}
a=-n,\qquad n=0,1,2,\ldots .
\end{equation}
Since \(a=\lambda-\varepsilon\), this gives
\begin{equation}
\varepsilon_n=n+\lambda,
\qquad
\varepsilon_n^2=(n+\lambda)^2 .
\label{app:eq_eps_scalar}
\end{equation}
The corresponding polynomial solution is proportional to a Gegenbauer
polynomial,
\begin{equation}
F_n(z)\propto
C_n^{(\lambda)}(\cos x),
\end{equation}
and therefore
\begin{equation}
u_n(x)=N_n(\sin x)^\lambda C_n^{(\lambda)}(\cos x).
\end{equation}

Using \(\lambda=\nu+1/2\) and
\begin{equation}
\varepsilon^2=\frac{\kappa^2}{2\Lambda}+\frac14 ,
\end{equation}
we find
\begin{equation}
\frac{\kappa_n^2}{2\Lambda}
=
\left(n+\nu+\frac12\right)^2-\frac14
=
(n+\nu)(n+\nu+1).
\end{equation}
Finally, since
\begin{equation}
\kappa^2=E^2-M^2-k_z^2 ,
\end{equation}
the scalar and longitudinal spectra are
\begin{equation}
E_n=
\pm
\sqrt{
M^2+k_z^2+2\Lambda(n+\nu)(n+\nu+1)
}.
\end{equation}

\subsection{Transverse sector}
\label{app:hypergeom_spectra_v}

The transverse modes satisfy the generalized trigonometric
Pöschl--Teller equation
\begin{equation}
\left[
-\frac{d^2}{dx^2}
+
\frac{\alpha_s^2-\frac14}{4\sin^2(x/2)}
+
\frac{\beta_s^2-\frac14}{4\cos^2(x/2)}
\right]u_s(x)
=
\varepsilon^2u_s(x),
\label{app:eq_PT_trans}
\end{equation}
where
\begin{equation}
\alpha_s=|b+s|,
\qquad
\beta_s=|b-s|.
\end{equation}
We again introduce
\begin{equation}
z=\sin^2\frac{x}{2}
=
\frac{1-\cos x}{2},
\qquad 0<z<1,
\end{equation}
and factor out the Friedrichs endpoint behaviour by writing
\begin{equation}
u_s(x)
=
\left(\sin\frac{x}{2}\right)^{\alpha_s+1/2}
\left(\cos\frac{x}{2}\right)^{\beta_s+1/2}
F_s(z).
\label{app:eq_ansatz_trans}
\end{equation}
Substituting Eq.~\eqref{app:eq_ansatz_trans} into
Eq.~\eqref{app:eq_PT_trans}, one obtains
\begin{equation}
\begin{split}
z(1-z)\frac{d^2F_s}{dz^2}
+
\left[
\alpha_s+1-(\alpha_s+\beta_s+2)z
\right]\frac{dF_s}{dz}
\\+
\left[
\varepsilon^2
-
\frac{(\alpha_s+\beta_s+1)^2}{4}
\right]F_s=0 .
\end{split}
\label{app:eq_hyper_trans}
\end{equation}
This is again the Gauss hypergeometric equation, with
\begin{equation}
c=\alpha_s+1,
\end{equation}
and
\begin{equation}
a_s=
\frac{\alpha_s+\beta_s+1}{2}-\varepsilon,
\qquad
b_s=
\frac{\alpha_s+\beta_s+1}{2}+\varepsilon .
\label{app:eq_ab_trans}
\end{equation}
The solution regular at \(z=0\) is therefore
\begin{equation}
F_s(z)={}_2F_1(a_s,b_s;c;z).
\end{equation}
The Friedrichs endpoint condition at $z=1$ is satisfied by the
polynomial solutions, obtained when the hypergeometric series
terminates. Imposing
\begin{equation}
a_s=-n,\qquad n=0,1,2,\ldots ,
\end{equation}
gives
\begin{equation}
\varepsilon_{s,n}
=
n+\frac{\alpha_s+\beta_s+1}{2},
\end{equation}
or
\begin{equation}
\varepsilon_{s,n}^2
=
\left(
n+\frac{\alpha_s+\beta_s+1}{2}
\right)^2 .
\label{app:eq_eps_trans}
\end{equation}
The corresponding polynomial solutions are Jacobi polynomials,
\begin{equation}
F_{s,n}(z)
\propto
P_n^{(\alpha_s,\beta_s)}(\cos x),
\end{equation}
so that
\begin{equation}
u_{s,n}(x)
=
N_{s,n}
\left(\sin\frac{x}{2}\right)^{\alpha_s+1/2}
\left(\cos\frac{x}{2}\right)^{\beta_s+1/2}
P_n^{(\alpha_s,\beta_s)}(\cos x).
\end{equation}

Using
\begin{equation}
\varepsilon^2=\frac{\kappa^2}{2\Lambda}+\frac14 ,
\end{equation}
we obtain
\begin{equation}
\frac{\kappa_{s,n}^2}{2\Lambda}
=
\left(
n+\frac{\alpha_s+\beta_s+1}{2}
\right)^2
-\frac14 .
\end{equation}
Therefore,
\begin{equation}
E_{s,n}
=
\pm
\sqrt{
M^2+k_z^2+
2\Lambda
\left[
\left(
n+\frac{\alpha_s+\beta_s+1}{2}
\right)^2
-\frac14
\right]
}.
\end{equation}

\section{Normalization of the longitudinal and transverse spin-1 modes}
\label{app:spin1_normalization}

In this appendix, we derive the radial normalization conditions for the
longitudinal and transverse modes of the projected spin-1 sector. We start from
the projected current
\begin{equation}
J^\mu
=
\frac{-i}{2M}
\left(
W_\lambda^\ast U^{\mu\lambda}
-
W_\lambda U^{\mu\lambda\ast}
\right),
\label{eq:app_projected_vector_current}
\end{equation}
where
\begin{equation}
U_{\mu\nu}=\nabla_\mu W_\nu-\nabla_\nu W_\mu .
\label{eq:app_Umunu}
\end{equation} Taking the temporal component of the projected current and using the orthonormal frame, we obtain
\begin{equation}
J^{0}
=
\frac{1}{M}\Im{
\left[
(W^{\hat{r}})^\ast U^{\hat{t}\hat{r}}+(W^{\hat{\phi}})^\ast U^{\hat{t}\hat{\phi}}+(W^{\hat{z}})^\ast U^{\hat{t}\hat{z}}
\right]},
\label{eq:app_projected_vector_current}
\end{equation}
where
\begin{align}
U^{\hat{t}\hat{r}} & =-U_{\hat{t}\hat{r}}=\partial_{t}W^{\hat{r}}+\partial_{r}W^{\hat{t}},\label{eq:app:Utr}\\
U^{\hat{t}\hat{\phi}} & =-U_{\hat{t}\hat{\phi}}=\partial_{t}W^{\hat{\phi}}+\frac{1}{S}\partial_{\phi}W^{\hat{t}},\label{eq:app:Utphi}\\
U^{\hat{t}\hat{z}} & =-U_{\hat{t}\hat{z}}=\partial_{t}W^{\hat{z}}+\partial_{z}W^{\hat{t}}.\label{eq:app:Utr}
\end{align}
The subsidiary condition $\nabla_{\mu}W^{\mu}=0$ gives
\begin{equation}
\partial_{t}W^{\hat{t}}+\partial_{r}W^{\hat{r}}+\frac{S^{\prime}}{S}W^{\hat{r}}+\frac{1}{S}\partial_{\phi}W^{\hat{\phi}}+\partial_{z}W^{\hat{z}}=0.
\label{eq:app:sub_eq}
\end{equation}
Considering the modal ansatz
\begin{equation}
W^{\hat{j}}(t,r,\phi,z)=e^{-iEt}e^{im\phi}e^{ik_z z}w^{\hat{j}}(r),
\label{eq:app:modal_ansatz}
\end{equation}
we find that
\begin{equation}
\begin{split}
J^{0} &= \frac{E}{M}\left( \vert w^{\hat{r}}\vert^{2}+\vert w^{\hat{\phi}}\vert^{2}+\vert w^{\hat{z}}\vert^{2} \right)
-\frac{1}{M}\Im\left[ (w^{\hat{r}})^{\ast}\partial_{r}w^{\hat{t}} \right]\\
&-\frac{m}{MS}\Re\left[ (w^{\hat{\phi}})^{\ast}w^{\hat{t}} \right]-\frac{k_{z}}{M}\Re\left[ (w^{\hat{z}})^{\ast}w^{\hat{t}} \right],
\end{split}
\label{eq:app:J0_geral}
\end{equation}
and Eq.~\eqref{eq:app:sub_eq} becomes
\begin{equation}
-iEw^{\hat{t}}+\partial_{r}w^{\hat{r}}+\frac{S^{\prime}}{S}w^{\hat{r}}+\frac{im}{S}w^{\hat{\phi}}+ik_{z}w^{\hat{z}}=0.
\label{eq:app:sub_eq2}
\end{equation}

\subsection{Longitudinal mode}
\label{app:longitudinal_normalization}

For the longitudinal mode, we consider
\begin{equation}
w^{\hat{r}}=w^{\hat{\phi}}=0,
\quad
w_{0}=w^{\hat{z}}\neq0.
\label{eq:app_longitudinal_ansatz}
\end{equation}
From the subsidiary condition \eqref{eq:app:sub_eq2}, we obtain the relation
\begin{equation}
w^{\hat{t}}=\frac{k_{z}}{E}w_{0}.
\label{eq:app:omegat}
\end{equation}
Substituting Eqs.~\eqref{eq:app_longitudinal_ansatz} and \eqref{eq:app:omegat} into Eq.~\eqref{eq:app:J0_geral}, one finds 
\begin{equation}
J^{0}_{(0)}=\frac{E^{2}-k_{z}^{2}}{ME}\vert w_{0}\vert^{2}.
\label{eq:app:J0_l}
\end{equation}
The normalization condition becomes
\begin{equation}
\int J^0_{(0)}\,d\tau
=
\frac{E^2-k_{z}^2}{ME}
\int{\vert w_{0}(r)\vert^{2}d\tau}=\pm 1.
\label{eq:app_norm_longitudinal_w}
\end{equation}
Using $d\tau=S(r)dr\,d\phi\,dz$, and after factoring out the trivial angular and longitudinal dependence, the admissibility of the longitudinal mode is determined by the radial condition
\begin{equation}
\int_0^{\pi/a}S(r)\vert w_{0}(r)\vert^{2}dr<\infty.
\label{eq:app_longitudinal_norm_finite}
\end{equation}

\subsection{Transverse modes}
\label{app:transverse_normalization}

We now turn to the transverse circular modes. For a pure circular-polarization mode we set
\begin{equation}
w_{-s}=0,
\quad
w_0=0,
\quad
w_{s}\neq 0,
\label{eq:app_transverse_ansatz}
\end{equation}
for $s=\pm 1$. Using the inverse relations of Eqs.~\eqref{eq:vector_helicity_basis1} and \eqref{eq:vector_helicity_basis2}, one obtains
\begin{equation}
w^{\hat r}=\frac{w_s}{\sqrt2},
\qquad
w^{\hat\phi}=-\frac{i s}{\sqrt2}\,w_s.
\label{eq:app_inverse_circular_components}
\end{equation}

The subsidiary condition now reads
\begin{equation}
w^{\hat{t}}
=
-\frac{i}{\sqrt2\,E}
\left(
\frac{d}{dr}
+\frac{S'+s m}{S}
\right)w_s.
\label{eq:app_wt_transverse}
\end{equation}
It is convenient to introduce
\begin{equation}
D_s\equiv \frac{d}{dr}+\frac{S'+s m}{S},
\qquad
\bar D_s\equiv \frac{d}{dr}-\frac{s m}{S},
\label{eq:app_Ds_defs}
\end{equation}
so that Eq.~\eqref{eq:app_wt_transverse} becomes
\begin{equation}
w^{\hat{t}}=-\frac{i}{\sqrt2\,E}D_s w_s.
\label{eq:app_wt_transverse_compact}
\end{equation}
Substituting Eqs.~\eqref{eq:app_transverse_ansatz}, \eqref{eq:app_inverse_circular_components}, and \eqref{eq:app_wt_transverse_compact} into Eq.~\eqref{eq:app:J0_geral}, one finds
\begin{equation}
J^0_{(s)}
=
\frac{E}{M}\vert w_s\vert^2
+\frac{1}{2ME}
\Re\!\left(
w_s^\ast\,\bar D_s D_s\,w_s
\right).
\label{eq:app_J0_transverse_Ds}
\end{equation}
The product \(\bar D_s D_s\) is given by
\begin{equation}
\bar D_s D_s
=
\frac{d^2}{dr^2}
+\frac{S'}{S}\frac{d}{dr}
+\frac{S''}{S}
-\frac{(m+sS')^2}{S^2}.
\label{eq:app_DbarD_identity}
\end{equation}
On the other hand, the radial equation for a pure circular-polarization mode is
\begin{equation}
\left[
\frac{d^2}{dr^2}
+\frac{S'}{S}\frac{d}{dr}
+\kappa^{2}
-\frac{(m+sS')^2}{S^2}
+\frac{S''}{S}
\right]w_s=0.
\label{eq:app_radial_transverse}
\end{equation}
Comparing Eqs.~\eqref{eq:app_DbarD_identity} and \eqref{eq:app_radial_transverse}, one obtains
\begin{equation}
\bar D_s D_s\,w_s
=-\kappa^{2}w_s=
-(E^2-k_{z}^2-M^2)\,w_s.
\label{eq:app_DbarD_on_ws}
\end{equation}
Substituting this result into Eq.~\eqref{eq:app_J0_transverse_Ds}, we arrive at
\begin{equation}
J^0_{(s)}
=
\frac{E^2+k_{z}^2+M^2}{2ME}\,
|w_s(r)|^2,
\qquad
s=\pm1.
\label{eq:app_J0_transverse_final}
\end{equation}
The normalization condition therefore reads
\begin{equation}
\int J^0_{(s)}\,d\tau
=
\frac{E^2+k_{z}^2+M^2}{2ME}
\int{\vert w_s(r)\vert^{2}d\tau}=\pm 1.
\label{eq:app_norm_transverse_w}
\end{equation}
Using $d\tau=S(r)dr\,d\phi\,dz$, and after factoring out the trivial angular and longitudinal dependence, the admissibility of the transverse mode is determined by the radial condition 
\begin{equation}
\int_0^{\pi/a}S(r)\,|w_s(r)|^2\,dr<\infty.
\label{eq:app_transverse_norm_finite}
\end{equation}

\bibliographystyle{spphys}       

\end{document}